\documentclass{article}
\usepackage[utf8]{inputenc}
\usepackage[utf8]{inputenc}
\usepackage[super]{nth}
\usepackage{mathtools}
\usepackage{url}
\usepackage{longtable}
\usepackage{colortbl}
\usepackage{graphicx,caption}
\usepackage{comment}
\usepackage[table,x11names,dvipsnames,table]{xcolor}
\usepackage{amsmath}
\usepackage{hyperref}
\usepackage[backend=bibtex]{biblatex}
\usepackage{authblk}
\usepackage{lineno}
\usepackage{float}

\usepackage[a4paper,
            bindingoffset=0.2in,
            left=1in,
            right=1in,
            top=1in,
            bottom=1in,
            footskip=.25in]{geometry}
\DeclareUnicodeCharacter{2212}{\ensuremath{-}}
\author[1]{Paul Triebkorn}
\author[1]{Huifang E. Wang}
\author[1]{Marmaduke Woodman}
\author[3,4]{Maxime Guye}
\author[1,2]{Fabrice Bartolomei}
\author[1]{Viktor Jirsa}

\affil[1]{Aix-Marseille Université, Institut National de la Santé et de la Recherche Médicale, Institut de Neurosciences des Systèmes (INS)UMR1106, Marseille 13005, France}
\affil[2]{APHM, Epileptology and Clinical Neurophysiology Department, Timone Hospital, Marseille 13005, France}
\affil[3]{Aix-Marseille Université, CNRS, CRMBM, Marseille 13005, France}
\affil[4]{APHM, Timone University Hospital, CEMEREM, Marseille 13005, France}

\newcommand{\beginsupplement}{%
        \setcounter{table}{0}
        \renewcommand{\thetable}{S\arabic{table}}%
        \setcounter{figure}{0}
        \renewcommand{\thefigure}{S\arabic{figure}}%
     }

\begin{document}
\title{Delay-Constrained Re-entry Governs Large-Scale Brain Seizures and Other Network Pathologies}
\maketitle
\begin{abstract}
Re-entry of travelling excitation loops is a long-suspected driver of human seizures, yet how such loops arise in patient brain networks—and how susceptible they are to targeted disruption—remains unclear. We reconstruct a millimetre-scale virtual brain from diffusion MRI of a drug-resistant epilepsy patient, embed excitable Epileptor neural fields, and show that realistic cortico-cortical delays are sufficient to generate self-sustaining re-entry. Systematic parameter sweeps reveal a narrow delay–coupling window that predicts oscillation frequency and seizure duration across 184 recorded seizures. Precisely timed biphasic stimuli or sub-millimetre virtual lesions abort re-entry in silico, yielding phase-dependent termination rules validated in intracranial recordings. Our framework exposes delay-constrained re-entry as a generic dynamical mechanism for large-scale brain synchrony and provides a patient-specific testbed for precision neuromodulation and minimally invasive disconnection.
\end{abstract}

\section{Introduction}
Epilepsy affects an estimated 46 million people worldwide and remains one of the most debilitating neurological disorders despite decades of pharmacological progress \cite{Beghi2019}. When seizures persist under medication, clinicians turn to resective surgery or neurostimulation, both of which demand millimetre-scale localisation of the epileptogenic zone (EZ) and a mechanistic understanding of how abnormal activity ignites, spreads and terminates \cite{Guery2021,Nascimento2023}. Yet even with invasive stereo-EEG, the spatiotemporal dynamics that link a focal trigger to cortex-wide seizures are only partially resolved.

Computational modelling promises to bridge this gap by providing a controllable, patient-specific replica of brain dynamics \cite{Depannemaecker2021}. The “virtual epileptic patient” (VEP) framework reconstructs each individual’s connectome from neuro-imaging, embeds a neural-mass model (the Epileptor) in every atlas parcel and infers regional excitability from recorded seizures \cite{Jirsa2017,Wang2023}. VEP has helped guide surgery in several centres, but its point-node architecture cannot capture intra-areal travelling waves or the sub-second phase lags that shape seizure propagation \cite{Martinet2017}. Recent extensions to spatially continuous neural fields reproduce local wave fronts on simplified 1-D or small cortical patches \cite{Proix2018,Sip2021}, yet they omit subject-specific long-range fibre anatomy that could enable network-scale feedback.

Re-entry—an excitation loop that revisits tissue after its refractory period—offers a unifying mechanism for such feedback. The phenomenon has been demonstrated in hippocampal slice cultures and chemically disinhibited rodent cortex, where recurrent excitatory connections sustain self-propagating waves \cite{sutula_unmasking_2007,lynch_recurrent_2000,christian_characteristics_1988,jin_enhanced_2006}. Similar cyclic patterns appear in in-vitro neuronal monolayers, disinhibited cat and rat cortex, and large-scale neural-network simulations that combine dense local and sparse long-range links \cite{Keren2016,Viventi2011,Huang2004,Jacob2019}. Human evidence is emerging: single-pulse stimulation of frontal epileptogenic cortex can provoke delayed or repetitive responses consistent with cortico-cortical loops \cite{Valentin2005,Goodfellow2012,wendling_identification_2001}. Together, these studies suggest that delay-mediated re-entry may couple microscopic excitability to macroscopic seizure spread, yet no model has tested this hypothesis in a full, patient-specific brain geometry.

Here we integrate sub-millimetre neural fields with high-resolution diffusion-MRI connectomes to create a personalised virtual brain that preserves both local cortical geometry and thousands of long-range fibre delays \cite{taylor_within_2017,naze_robustness_2021}. Each vertex hosts the Epileptor in an excitable regime, allowing travelling waves to emerge spontaneously or in response to stimulation. Using this platform we map the delay–coupling parameter space and identify a narrow corridor that sustains large-scale re-entry consistent with intracranial recordings; we quantify seizure propagation and termination as functions of loop length, conduction speed and refractoriness; and we test targeted interventions—biphasic stimulation, virtual fibre transection and multiple subpial cuts—to determine how, when and where re-entry can be interrupted.

By converging patient anatomy, neural-field theory and systematic perturbation, we ask whether delay-constrained re-entry is (i) sufficient to explain whole-brain seizure dynamics and (ii) vulnerable to clinically feasible disruptions. Our answers, we argue, move virtual-brain modelling from descriptive replay to predictive neuromodulation design and lay the groundwork for precision therapies in drug-resistant epilepsy.

\section{Results}

\subsection{High resolution models and its application in epilepsy}

\begin{figure}[H]
\includegraphics[width=\textwidth]{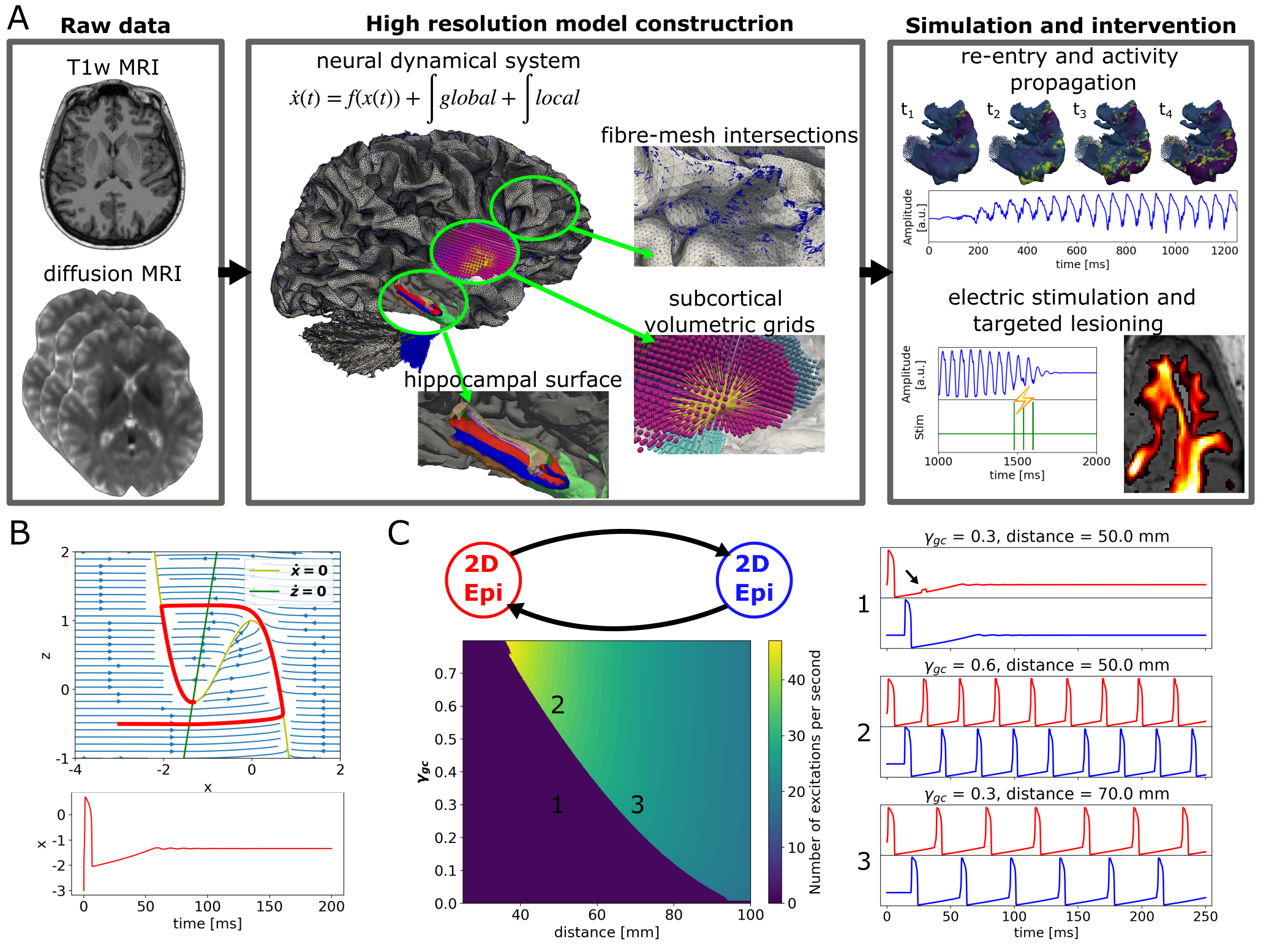}
\centering
\caption{High-resolution virtual brain model reconstruction for delay-constrained re-entry based seizure simulations. \textbf{(A)} T1w and diffusion MRI together with tractography are used to reconstruct the cortical surface and to estimate connections between points of the surface (left). Subcortical areas are either represented by volumetric grids or surfaces (middle). The resulting connectomes for local and global connections are used together with a dynamical model, called the 2D Epileptor, to simulate epileptic brain dynamics. The neural activity at each point of the network is given by the dynamical equations of the 2D Epileptor together with the sum of global and local connections. The 2D Epileptor is parameterized to an excitable regime to test the hypothesis of re-entry mechanisms in epileptic seizures (right). Subsequently, the model is used to investigate electrical stimulation and targeted lesioning to prevent seizures. \textbf{(B)} The phase space of a single 2D Epileptor in an excitable regime is displayed. Yellow and green lines indicate nullclines. The time series of the red example trajectory is shown below. \textbf{(C)} Re-entry excitation is examined in two delay-coupled 2D Epileptors as a function of coupling strength $\gamma_{gc}$ and inter-node distance. Only the 1st Epileptor (red time series) is initialized beyond the stable focus, generating the first excitation. The 2nd Epileptor (blue time series) is initialized on the stable focus. Simulations are run for 1s and the number of excitations of the 2$^{nd}$ Epileptor are counted. Example time series of both 2D Epileptors at numbered positions in parameter space are shown on the right. For point 1 in parameter space no re-entry can be observed, as the perturbation of the 2$^{nd}$ Epileptor onto the 1$^{st}$ Epileptor is not sufficiently strong or delayed enough (black arrow). Increasing the coupling strength or distance leads to re-entry at points 2 and 3 in parameter space, with different re-entry rate. }
\label{fig:workflow_overview}
\end{figure}

Using structural T1-weighted MRI and diffusion-weighted tractography of an exemplar patient, we reconstructed a millimetre-scale virtual brain to probe re-entry mechanisms and candidate interventions (Fig. \ref{fig:workflow_overview}A). The cortical ribbon and those subcortical structures that can be unfolded onto a surface—the hippocampus and cerebellum—were tessellated as triangulated meshes, whereas nuclei such as thalamus and amygdala were represented by volumetric point grids.

\begin{figure}[h]
\includegraphics[width=\textwidth]{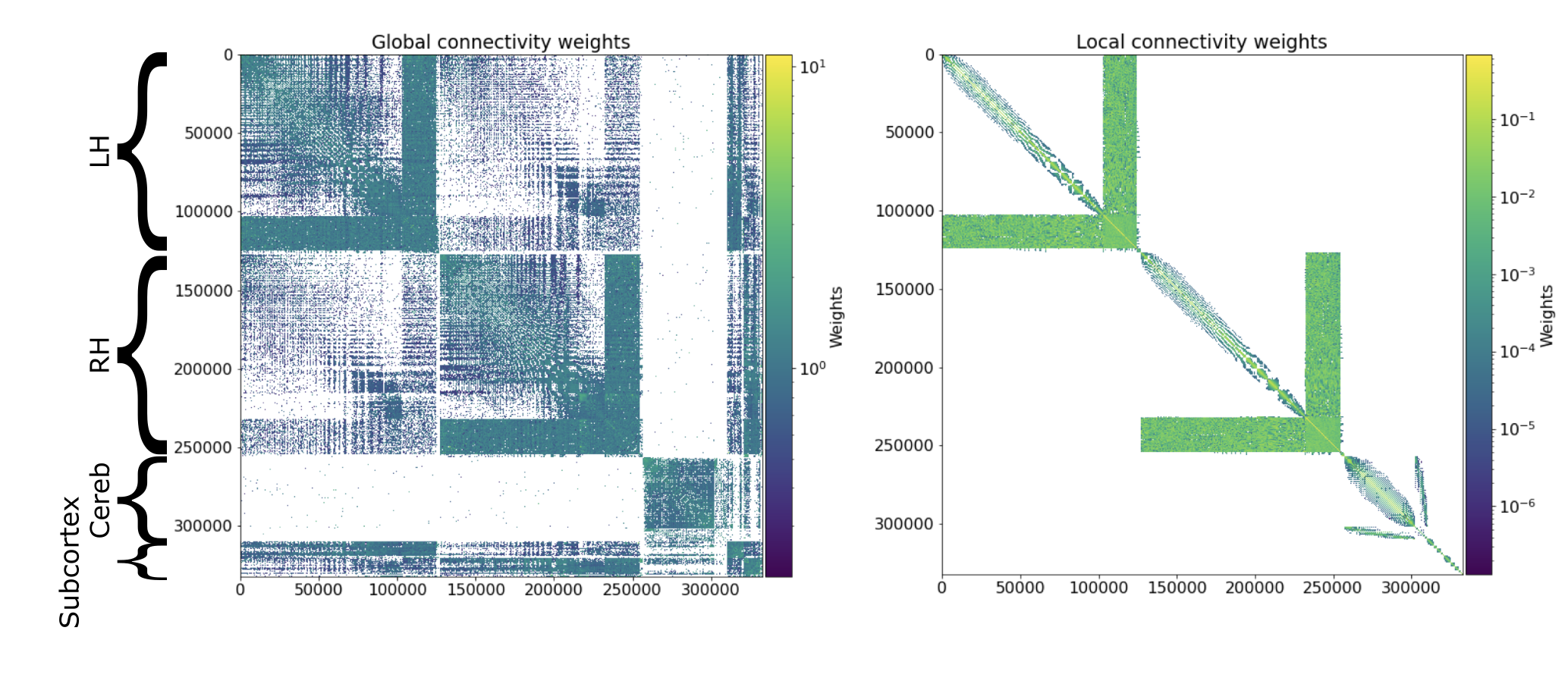}
\centering
\captionsetup{labelformat=empty}
\caption{Figure S1 Global and local connectivity matrix for the high-resolution brain model.}
\label{fig:suppl_global_local_connectivity}
Global and local connections are estimated between all 332455 vertices of the model which results in 2 connectivity matrices. The block-like structure of each matrix is given by the anatomical seperation into left- (LH) and right-hemisphere (RH), as well as cerebellum (Cereb) and subcortical structures. The vertex numbering follows the Freesurfer based surface generation process.
\end{figure}

Local synaptic coupling between neighbouring gray-matter vertices decays exponentially with geodesic (surface) or Euclidean (volume) distance, while long-range axonal coupling is derived from streamline intersections with surface or volume meshes; the resulting vertex-to-vertex matrices are shown in Fig. \ref{fig:suppl_global_local_connectivity}.  Each vertex hosts a 2D Epileptor in an excitable regime: while an isolated node returns to its fixed point after perturbation, delay-coupling via the white-matter network enables sustained circulating oscillations. Known as re-entry, these oscillations are excitation waves that circumnavigate an excitable network and revisit their point of origin after a local refractory period, facilitated by recurrent connectivity and finite axonal delays. To illustrate the mechanics, we first analysed a toy system of two bidirectionally delay-coupled 2D Epileptors. Each unit was tuned slightly below the super-critical Hopf bifurcation ($x_0$ = –1.2916), producing a stable focus in the phase plane (Fig. \ref{fig:workflow_overview}B). A modest perturbation of the first oscillator triggered a single large excursion, whereas the second remained at rest. We then swept coupling strength  $\gamma_{gc}$ and distance $d_{i,j}$ — the latter sets the propagation delay — to map conditions that sustain cycling of activity. With weak coupling and short delays, the return excitation arrives during refractoriness and activity extinguishes (example 1 in Fig. \ref{fig:workflow_overview}C). Increasing either  $\gamma_{gc}$ or $d_{i,j}$ pushes the downstream oscillator across threshold after it has recovered, creating a self-maintained ping-pong loop; larger separations lengthen the inter-spike interval (examples 2 and 3 in Fig. \ref{fig:workflow_overview}C) .

This platform allows us, first, to chart the parameter space that governs the emergence, persistence, and self-limitation of re-entry and, second, to test two families of interventions—phase-locked electrical stimulation and targeted surgical disconnection—that aim to extinguish ongoing loops by manipulating local or global connectivity. 

\subsection{Modeling re-entry excitation}
\label{subsection:modelling}
\begin{figure}[H]
\includegraphics[width=\textwidth]{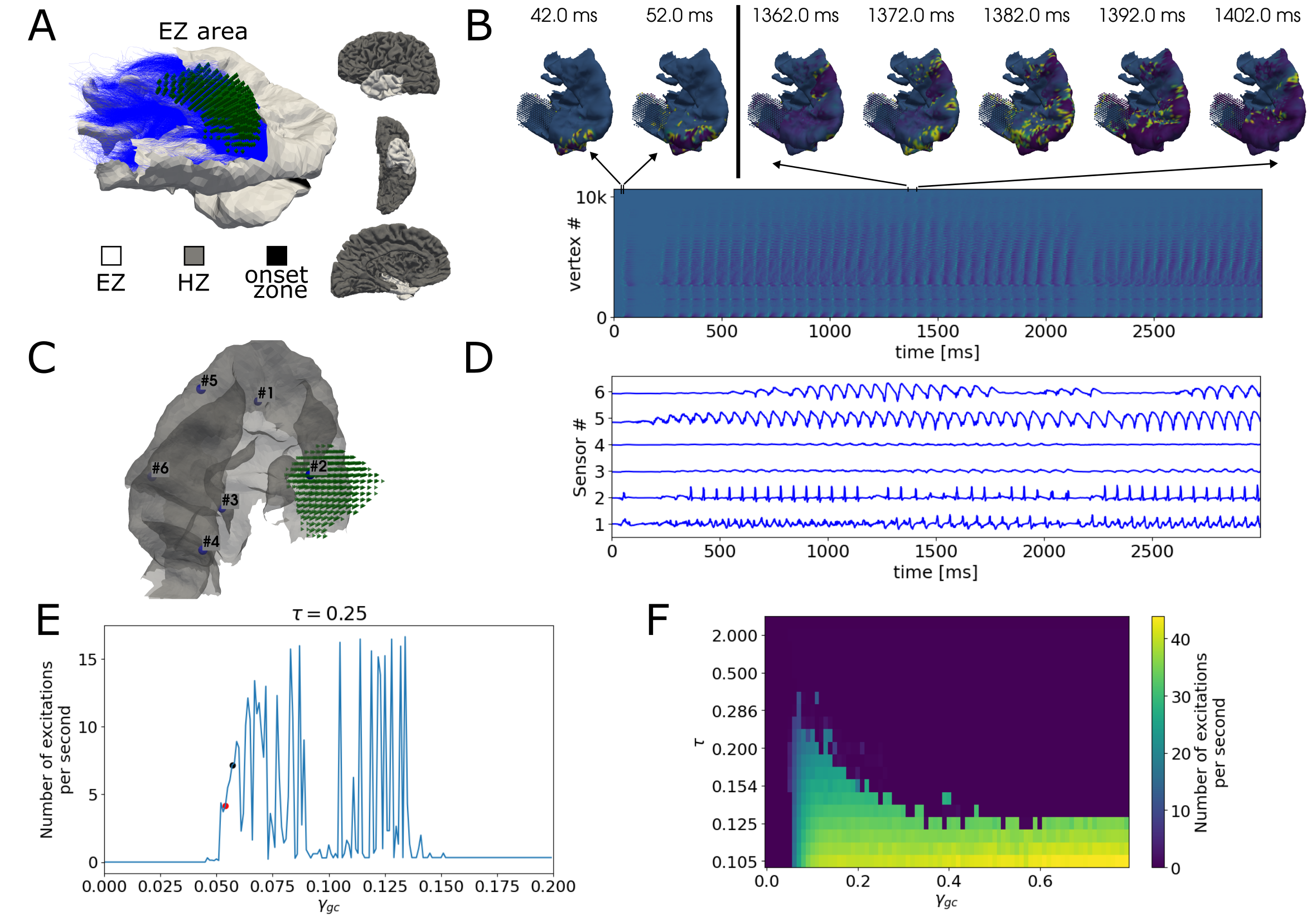}
\centering
\caption{Modeling re-entry excitation on a patch of the left temporal lobe.}
\label{fig:re-entry_temporal_lobe}
\textbf{(A)} Mesial view of the patch of cortex on the anterior temporal lobe that was used for simulations. White tissue is modeled as EZ. The black patch is the onset zone, from where activity starts to propagate. The rest of the cortex is colored in gray, with very low excitability, not being involved in the seizure activity. Green tetrahedra represent the volumetric grid of the Amygdala. White matter tracts are displayed in blue. \textbf{(B)} Example simulation of re-entry on the cortical patch. At the top, snapshots of neural activity on the cortical patch are displayed at indicated time points. Excitations can be seen as yellow waves traveling across the surface. Below, a space-time plot, shows re-entrant neural activity of the full simulation. \textbf{(C)} Superior view of the patch of cortex showing the location and number of implanted virtual contacts. \textbf{(D)} Corresponding recorded electrical signal on the contacts of (C) given the simulation in (B). \textbf{(E)} Parameter exploration across coupling strength $\gamma_{gc}$ from 0 to 0.2, with $\tau=0.25$. The variability of re-entry rates is rich in this parameter range. Three seconds of neural activity on the cortical patch were simulated across the indicated parameter space. The y-axis indicates the number of excitations per second averaged across the tissue. The black and red dot indicate the parameter for the simulations in (B) and Fig. (\ref{fig:suppl_self_limiting_re-entry}), respectively. \textbf{(F)} Parameter exploration across coupling strength $\gamma_{gc}$ and time scale $\tau$ of the Epileptor. The color bar indicates the number of excitations per second averaged across the tissue, for three-second simulation length.
\end{figure}

Moving beyond the minimal pair of two delay-coupled 2D Epileptors, we placed the same excitable Epileptor on a patient-specific patch of left anterior temporal cortex extracted with the VEP atlas \cite{Wang2021} (Temporal Pole through Rhinal cortex, plus amygdala) (Fig. \ref{fig:re-entry_temporal_lobe}A). To limit computation we treated the remainder of the brain as quiescent and simulated only this patch plus its long-range fibres (Fig. \ref{fig:re-entry_temporal_lobe}B). Local $\gamma_{lc}$  and global $\gamma_{gc}$ coupling were scaled in a fixed 1:1 ratio, reflecting anatomical estimates\cite{braitenberg_cortex_1998}. All vertices were initialised at the stable focus except those in the designated onset zone, which received a small offset to seed a single pulse.

Snapshots from a three-second simulation (Fig. \ref{fig:re-entry_temporal_lobe}B) show activity leaving the onset zone as concentric surface waves while simultaneously “jumping” via white-matter shortcuts. A corresponding video of the simulation is given in Supplementary V1. The space-time raster reveals periodic re-activation of distant vertices, diagnostic of large-scale re-entry, and six virtual SEEG contacts reproduce the anterior-to-posterior spread observed clinically (Fig. \ref{fig:re-entry_temporal_lobe}C,D). Systematic variation of  $\gamma_{gc}$ identified three regimes (Fig. \ref{fig:re-entry_temporal_lobe}E):
$\gamma_{gc} < 0.05 $: perturbations fail to propagate and die out;
$0.05 < \gamma_{gc} < 0.15 $: waves circulate continuously or as self-limiting transients;
$\gamma_{gc} > 0.15 $: excitation outruns recovery, recruiting the entire patch at once and imposing global refractoriness.

Because axonal delays are fixed by anatomy, we next modulated the intrinsic time-scale $\tau$ of the Epileptor (Fig. \ref{fig:re-entry_temporal_lobe}F). High $\tau > 0.33$ lengthens refractory periods so much that no coupling strength can support re-entry; intermediate $0.125<\tau<0.33$ permits loops across a window of $\gamma_{gc}$; and very low $\tau < 0.125$ raises excitability enough that re-entry emerges above a sharp $\gamma_{gc}$ threshold .

\begin{figure}[H]
\includegraphics[width=\textwidth]{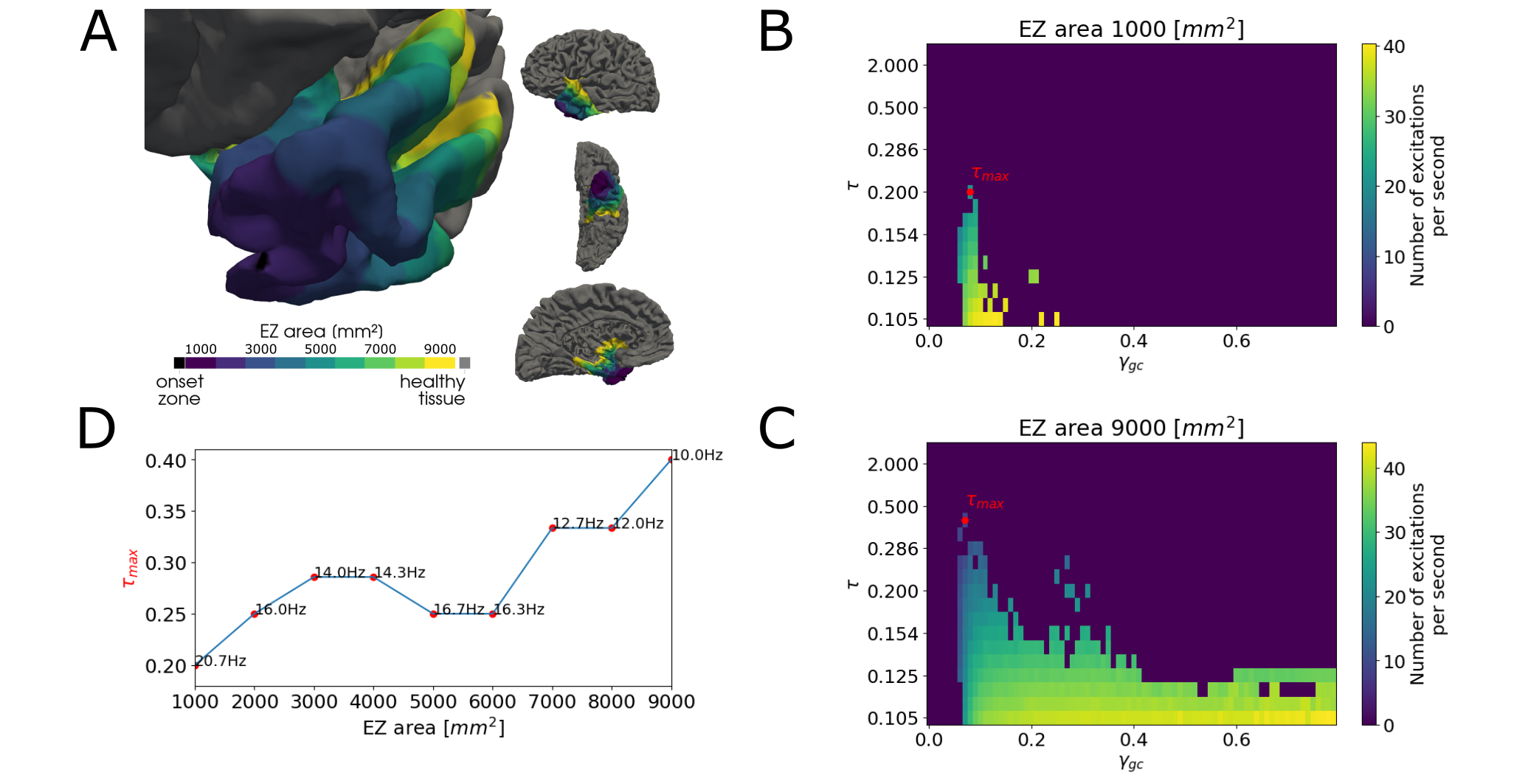}
\centering
\caption{Dependency of re-entry excitation on the size of the EZ.}
\label{fig:re-entry_EZ_size}
\textbf{(A)} Cortical surface showing the size of the EZ used for simulation. Color indicates the extent of the different EZ, with smaller EZs being subsets of larger ones. \textbf{(B)} Parameter exploration across coupling strength $\gamma_{gc}$ and time scale $\tau$ of the Epileptor for an EZ of size 1000 $mm^2$. For each parameter combination the average number of excitations per second is plotted for a three second simulation. The red dot indicates $\tau_{max}$ for which re-entry becomes possible for the given EZ size. 
\textbf{(C)} Same as (B) but for EZ size 9000 $mm^2$. \textbf{(D)} The $\tau_{max}$, identified from the parameter explorations, as a function of the EZ size, together with the corresponding frequency of the re-entry dynamics. 
\end{figure}

Finally, we asked how the extent of the EZ constrains feasible $\tau$. Growing the EZ from 1,000 mm$^2$ to 9,000 mm$^2$ increased the maximum $\tau$ (i.e. $\tau_{max}$) that still allows re-entry from 0.2 to 0.4 (Fig. \ref{fig:re-entry_EZ_size}A–D). Larger zones therefore support slower, lower-frequency loops because their intrinsic delays are longer. In all cases $\tau_{max}$ rose monotonically with EZ area (Fig. \ref{fig:re-entry_EZ_size}D).

Together, these experiments establish the chain of our mechanistic reasoning: minimal two-node loops demonstrate the fundamental delay–gain principle; embedding the same dynamics in patient geometry shows how local waves and tract-mediated jumps cooperate; and scaling analyses delineate the specific combinations of $\gamma_{gc}$, $\tau$, and EZ size that permit, limit, or extinguish re-entry in realistic tissue.

\subsection{The effect of stimulation on re-entry}

Having verified that the excitable cortical network sustains large-scale re-entry, we next asked whether a brief, local perturbation could extinguish the loop. The guiding idea is classical: if a stimulus excites a strategically chosen patch of cortex and immediately drives it into refractoriness, any incoming activity that returns via the long-range network will arrive during that refractory window and die out.

\begin{figure}[H]
\includegraphics[width=\textwidth]{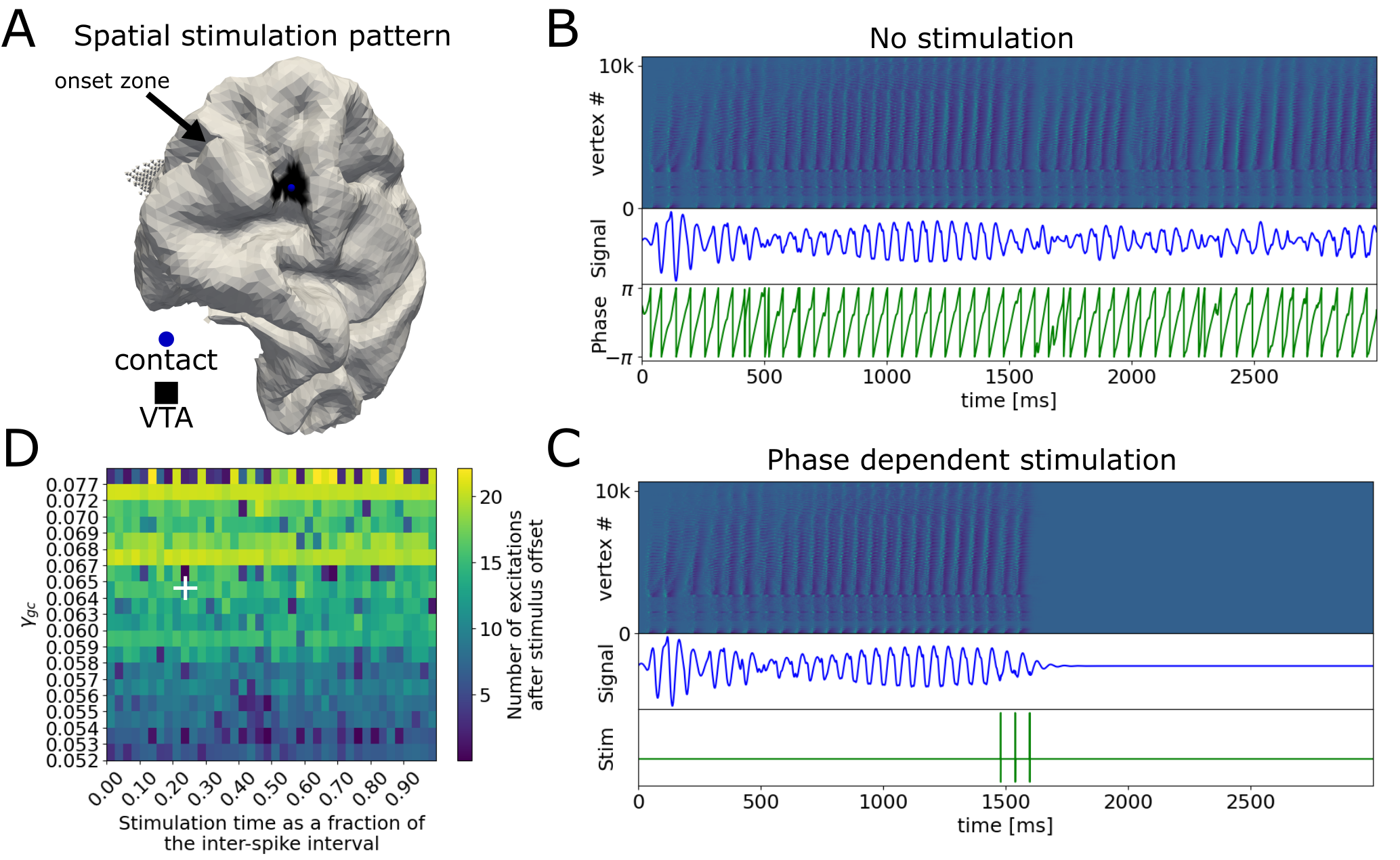}
\centering
\caption{Electrical stimulation to counteract re-entry excitation.}
\label{fig:electrical_stimulation}
\textbf{(A)} Inferior view of the cortical patch EZ used for simulation. The blue sphere indicates the contact used for sensing neural activity and applying an electrical stimulus. The black patch on the cortical surface represents the VTA, that gets excited by a stimulus of the contact. \textbf{(B)} Space-time plot with re-entry activity of an example simulation. Below, the filtered signal as recorded on the contact and the corresponding instantaneous phase. \textbf{(C)} Same as (B), but in the lower panel showing the time course of the 3-pulse stimulation applied to the simulation, successfully ending re-entry dynamics. \textbf{(D)} Parameter space exploration across coupling scaling $\gamma_{gc}$ and stimulation within the inter-spike interval of 2 subsequent re-entries. The color indicates the average number of excitations per second across the cortical patch after stimulus offset. The white cross indicates the parameter combination for the simulation in (C).
\end{figure}

We inserted one SEEG-like depth contact into the anterior temporal lobe, a location accessible in clinical practice (Fig. \ref{fig:electrical_stimulation}A). Neural activity is projected to the contact with the electromagnetic forward model, and the same electrode delivers stimulation. To mimic charge-balanced clinical pulses we used a biphasic waveform: 210 $\mu$s cathodic phase, followed by an anodic phase twice as long at half amplitude, a design shown to excite tissue efficiently while limiting electrochemical damage \cite{merrill_electrical_2005}. Finite-element estimates place the resulting spherical volume of tissue activated (VTA) at ~150 $mm^3$, well within published ranges \cite{butson_role_2006}.

Re-entry dynamics were simulated for 3 s, the contact signal was Hilbert-transformed, and the instantaneous phase of the last full inter-spike interval (ISI) before t = 1.5 s was extracted (Fig. \ref{fig:electrical_stimulation}B). Three identical pulses were then delivered, beginning at a manually-selected phase expressed as a percentage of the ISI; successive pulses were separated by exactly one ISI (Fig. \ref{fig:electrical_stimulation}C).

\begin{figure}[h]
\includegraphics[width=\textwidth]{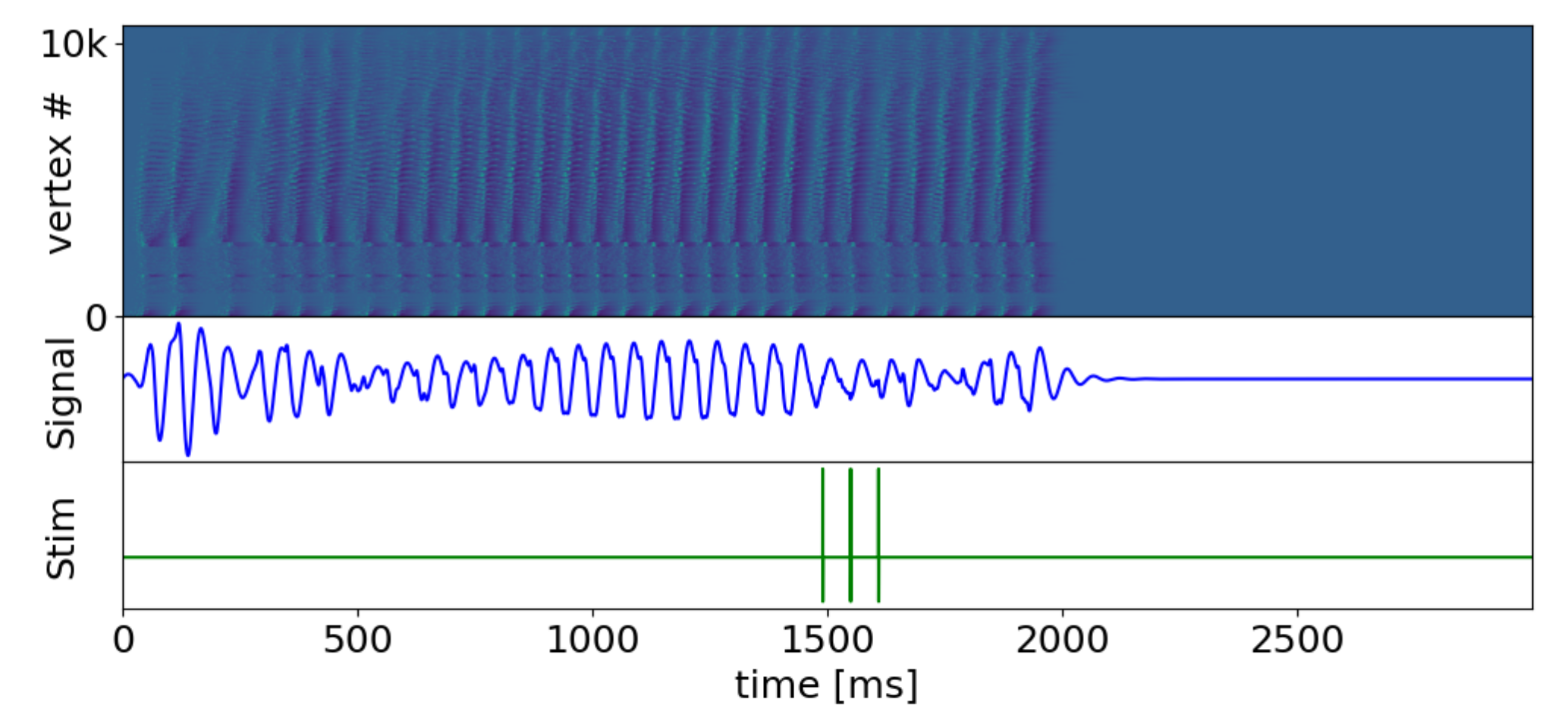}
\centering
\captionsetup{labelformat=empty}
\caption{Figure S3 Stimulation leads to end of re-entry within a few cycles after stimulation offset.}
\label{fig:suppl_stim2}
Space-time plot with re-entry activity of an example simulation. Below, the filtered signal as recorded on the contact and the corresponding instantaneous phase.
\end{figure}

\begin{figure}[h]
\includegraphics[width=\textwidth]{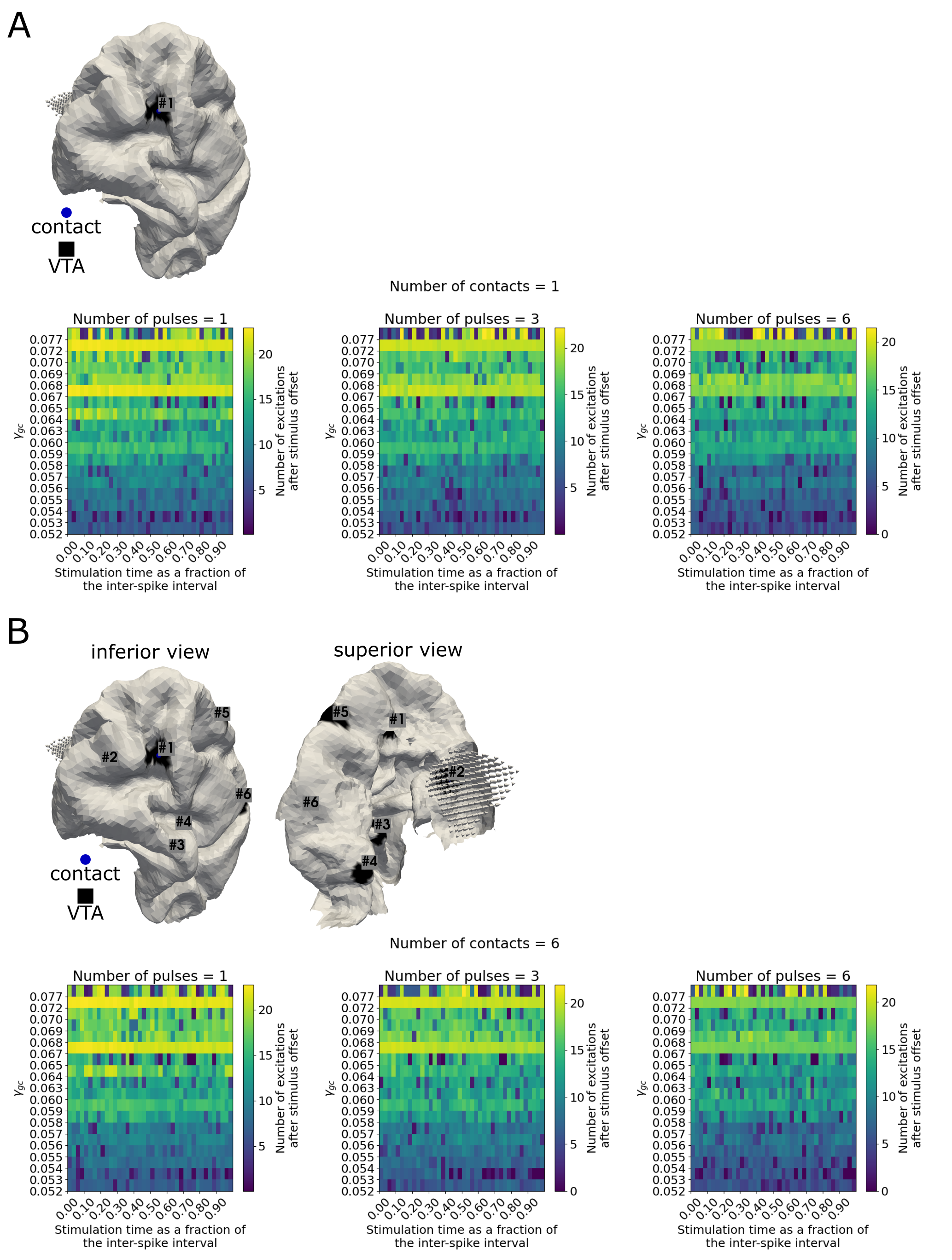}
\centering
\captionsetup{labelformat=empty}
\caption{Figure S4 Stimulation with one, three or six pulses and one or six dsitributed contacts.}
\label{fig:suppl_stim1}
\textbf{(A)} Inferior view of the cortical patch EZ used for simulation. The blue sphere indicates the contact used for sensing neural activity and applying an electrical stimulus. The black patch on the cortical surface represents the volume of tissue activated (VTA), that gets excitated by a stimulus of the contact. Below the parameter space exploration for stimulation success across parameters coupling strength $\gamma_{gc}$ and stimulation onset time within the inter spike interval. The color indicates the average number of excitations per second across the cortical patch after stimulus offset. The exploration was done for a one-, three- and six-pulses stimulus respectively.
\textbf{(B)} Same as (A) but for 6 spatially distributed contacts.
\end{figure}

Across the coupling range $0.052<\gamma_{gc}<0.08$ —where re-entry is otherwise self-sustaining—the three-pulse protocol abolished oscillations only when triggered inside narrow phase windows (Fig. \ref{fig:electrical_stimulation}D). Outside those windows the stimulus either failed outright or delayed termination by a few cycles. Supplemental sweeps confirmed that varying pulse count (1, 3 or 6) or distributing electrodes across six contacts produced qualitatively identical phase-selective success curves (Fig. \ref{fig:suppl_stim2} -  \ref{fig:suppl_stim1}).

In sum, a single depth electrode delivering three well-timed biphasic pulses can reliably halt re-entrant seizures in silico, but efficacy hinges sharply on the phase of delivery— an observation that directly informs closed-loop neuromodulation algorithms.

\subsection{The effect of connectivity manipulation on re-entry}

To probe surgical strategies that act not through stimulation but through structural disconnection, we removed selected edges from the patient’s connectome and measured the change in re-entry burden (mean number of excitations per second over the EZ).

\begin{figure}[H]
\includegraphics[width=\textwidth]{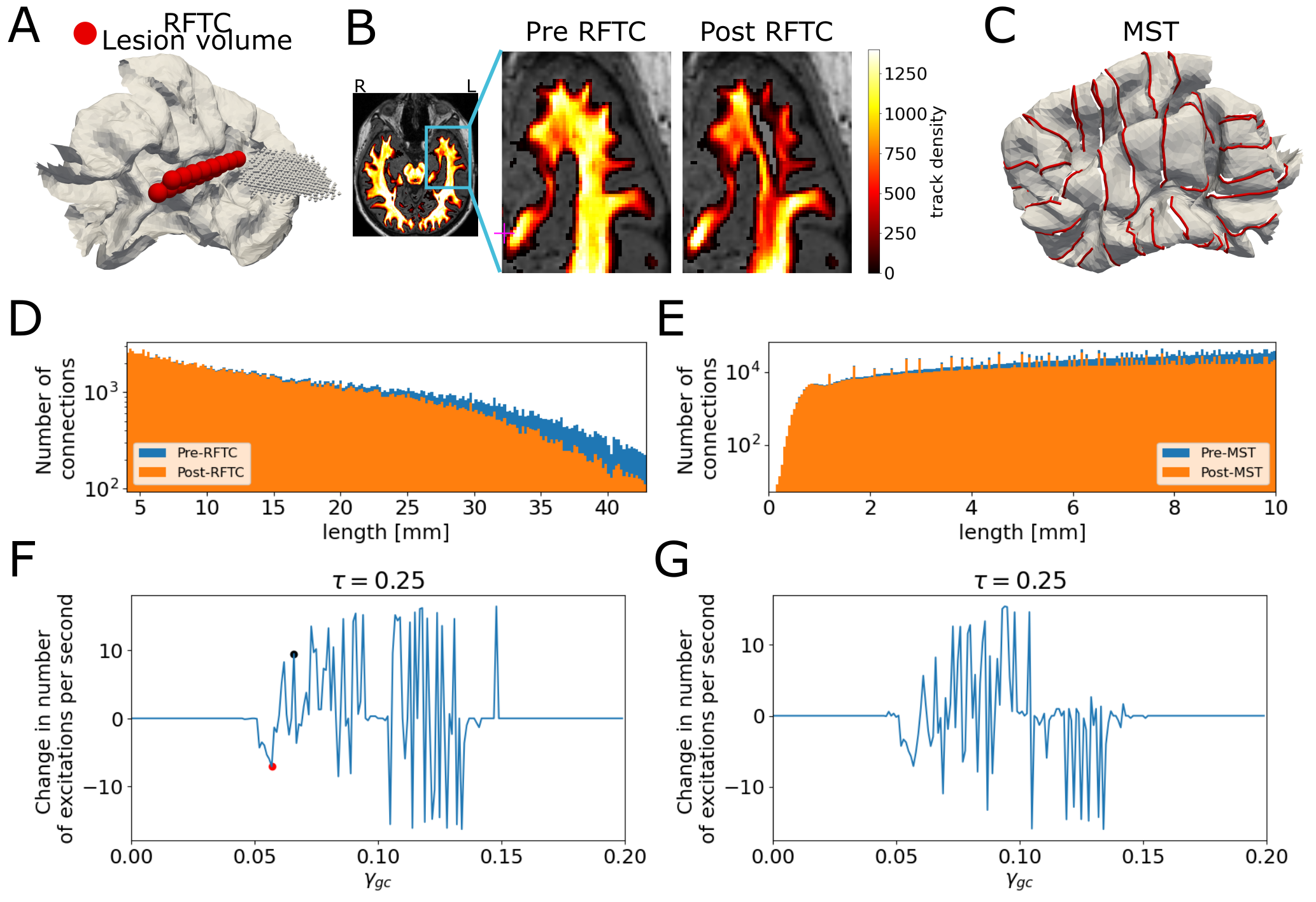}
\centering
\caption{Probing connectivity lesion effects on re-entry.}
\label{fig:white_matter_lesioning}
\textbf{(A)} Mesial view of the cortical patch that was used for simulations. The red spherical volumes (each radius = 2mm) indicate the white matter tissue that gets lesioned by RFTC. \textbf{(B)} Track density image of the left temporal lobe white matter prior and after introducing the lesions specified in (A). \textbf{(C)} Trajectories (red lines) of MST running parallel across all gyri of the EZ approximately every 5mm. Selected triangles were removed from the surface mesh, resulting in holes and local disconnections between vertices.
\textbf{(D)} Histogram of long-range connections in the left anterior temporal lobe prior and after RFTC lesioning, sorted by their length. \textbf{(E)} Histogram of short-range connections in the left anterior temporal lobe prior and after MST lesioning, sorted by their length. 
\textbf{(F)} Parameter exploration across coupling strength $\gamma_{gc}$. Three seconds of neural activity on the cortical patch were simulated across the indicated parameter space. The y-axis indicates the change from pre- to post-lesional average number of excitations per second across the tissue. The red and black dot indicate the parameter for the simulations in Fig. \ref{fig:suppl_virt_surgery_ts}, showing the destabilizing and stabilizing effect on re-entry. \textbf{(G)} same as (F) but for MST.
\end{figure}

\begin{figure}[h]
\includegraphics[width=\textwidth]{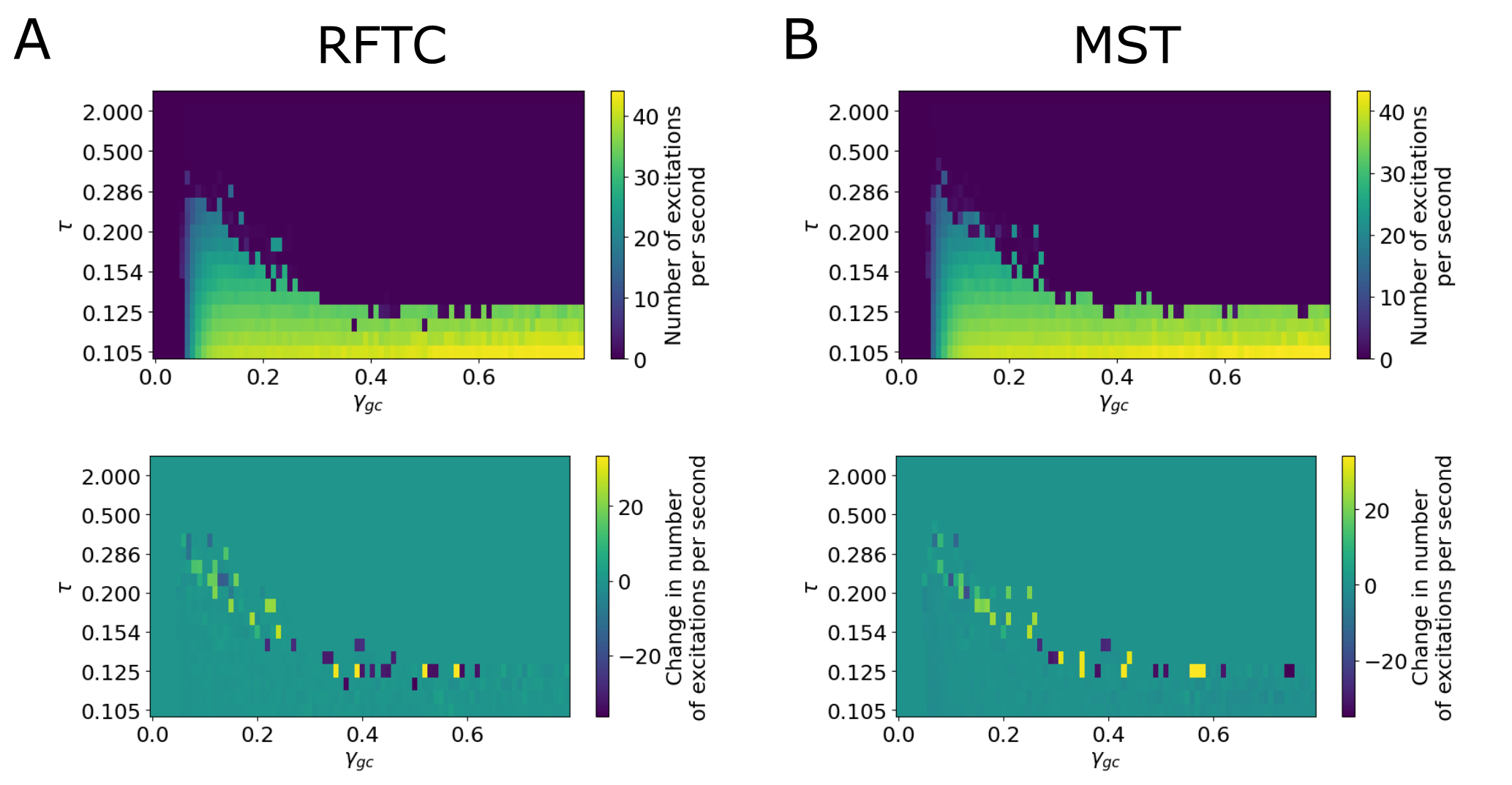}
\centering
\captionsetup{labelformat=empty}
\caption{Figure S5 Parameter exploration after RFTC and MST.}
\label{fig:suppl_virt_surgery}
(A) Top panel shows the parameter exploration across coupling strength $\gamma_{gc}$ and time scale $\tau$ of the Epileptor. For each parameter combination the average number of excitations per second is plotted for a three second simulation, after applying radio frequency thermocoagulation (RFTC). Lower panel shows the difference of the parameter exploration between pre- and post-RFTC. (B) Same as (A) but for multiple subpial transsection.
\end{figure}

\begin{figure}[h]
\includegraphics[width=\textwidth]{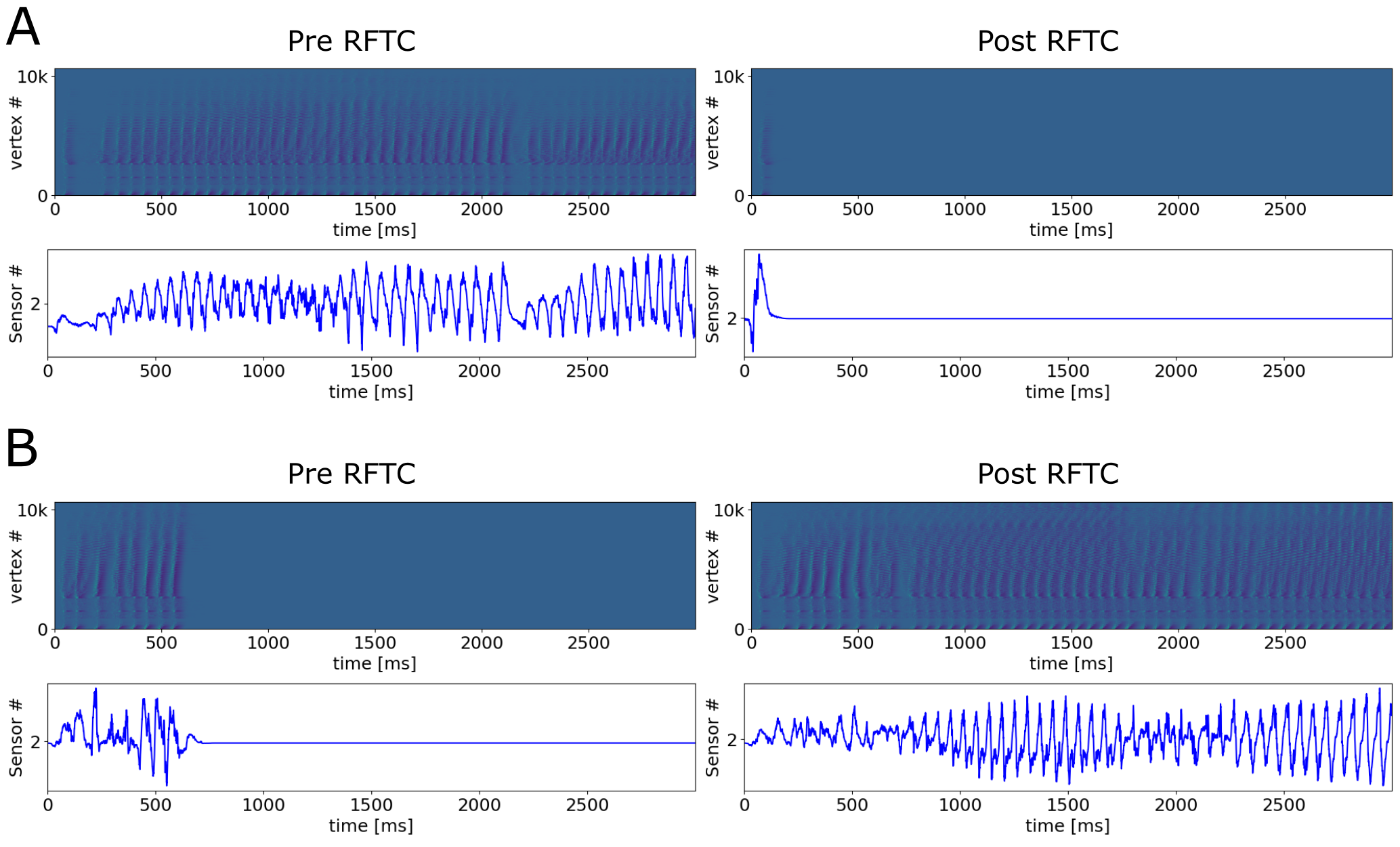}
\centering
\captionsetup{labelformat=empty}
\caption{Figure S6 Example time series for the effect of RFTC.}
\label{fig:suppl_virt_surgery_ts}
(A)  Left and right panel show the pre- and post-RFTC space-time plots of the simulations with $\gamma_{gc}=0.057$, respectively. Below the projection to a SEEG contact from Fig. \ref{fig:re-entry_temporal_lobe}.  The pre-RFTC simulation shows sustained re-entry, highlighted by the repeating waves of activity. The RFTC lesions however stabilises the system, preventing re-entry. (B) Same as (A) but with $\gamma_{gc}=0.061$. Showing the possible de-stabilising effect of RFTC lesioning.
\end{figure}

Monopolar Radio-frequency thermocoagulation (RFTC) of long-range fibres was modelled as 2 mm-radius spherical lesions around a contact (empirical range is 1.5–3 mm \cite{bourdillon_stereo-electro-encephalographyguided_2016}) along two virtual SEEG electrodes (8 contacts, 2 mm contact length, 1.5 mm spacing) which were placed in the left anterior temporal white matter, mimicking clinically feasible trajectories (Fig. \ref{fig:white_matter_lesioning}A). Every streamline intersecting a sphere was deleted; short-range U-fibres were largely spared, whereas long corticocortical projections were disproportionately affected (Fig. \ref{fig:white_matter_lesioning}B,D). Across coupling strengths in the range $0<\gamma_{gc}< 0.2$ the intervention produced a non-monotonic effect (Fig. \ref{fig:white_matter_lesioning}F): at low $\gamma_{gc}$ the lesion stabilised the network, truncating loops that were otherwise sustained; near the lower edge of the re-entry corridor it destabilised the system, creating loops where none existed; at high $\gamma_{gc}$ it had negligible impact because activity already failed through global refractoriness.
The duality arises because excising a tract both lengthens effective delay and weakens feedback; which influence dominates depends on baseline $\gamma_{gc}$ and $\tau$ (Fig. \ref{fig:suppl_virt_surgery}A). Two illustrative simulations — one stabilising, one destabilising — are shown in Fig. \ref{fig:suppl_virt_surgery_ts}.

We next simulated Multiple subpial transection (MST), a technique that severs tangential intracortical axons while preserving long-range tracts \cite{morrell_multiple_1989}. Parallel cuts, 5 mm apart, were laid across the anterior temporal gyri (Fig. \ref{fig:white_matter_lesioning} C). Only local edges shorter than 10 mm were removed, sharply reducing peak degree within the EZ but leaving deep-white-matter connectivity unchanged (Fig. \ref{fig:white_matter_lesioning} E). Parameter sweeps again revealed a bidirectional outcome (Fig. \ref{fig:white_matter_lesioning} G): moderate $\gamma_{gc}$  values were most sensitive—some converted sustained loops into transient flickers, others unmasked new loops—whereas extreme $\gamma_{gc}$  values were little affected. Notably, MST shifted the upper $\gamma_{gc}$  boundary of the re-entry corridor downward, implying that weakening local recurrence can substitute for stronger global delays (Fig. \ref{fig:suppl_virt_surgery}B).

Taken together, these in-silico dissections show that surgical success depends on matching the lesion type to the patient’s specific delay–coupling operating point — a principle that argues for the necessity of individual virtual brain testing before real-world intervention.

\subsection{Re-entry activity invading a propagation zone}

\begin{figure}[H]
\includegraphics[width=\textwidth]{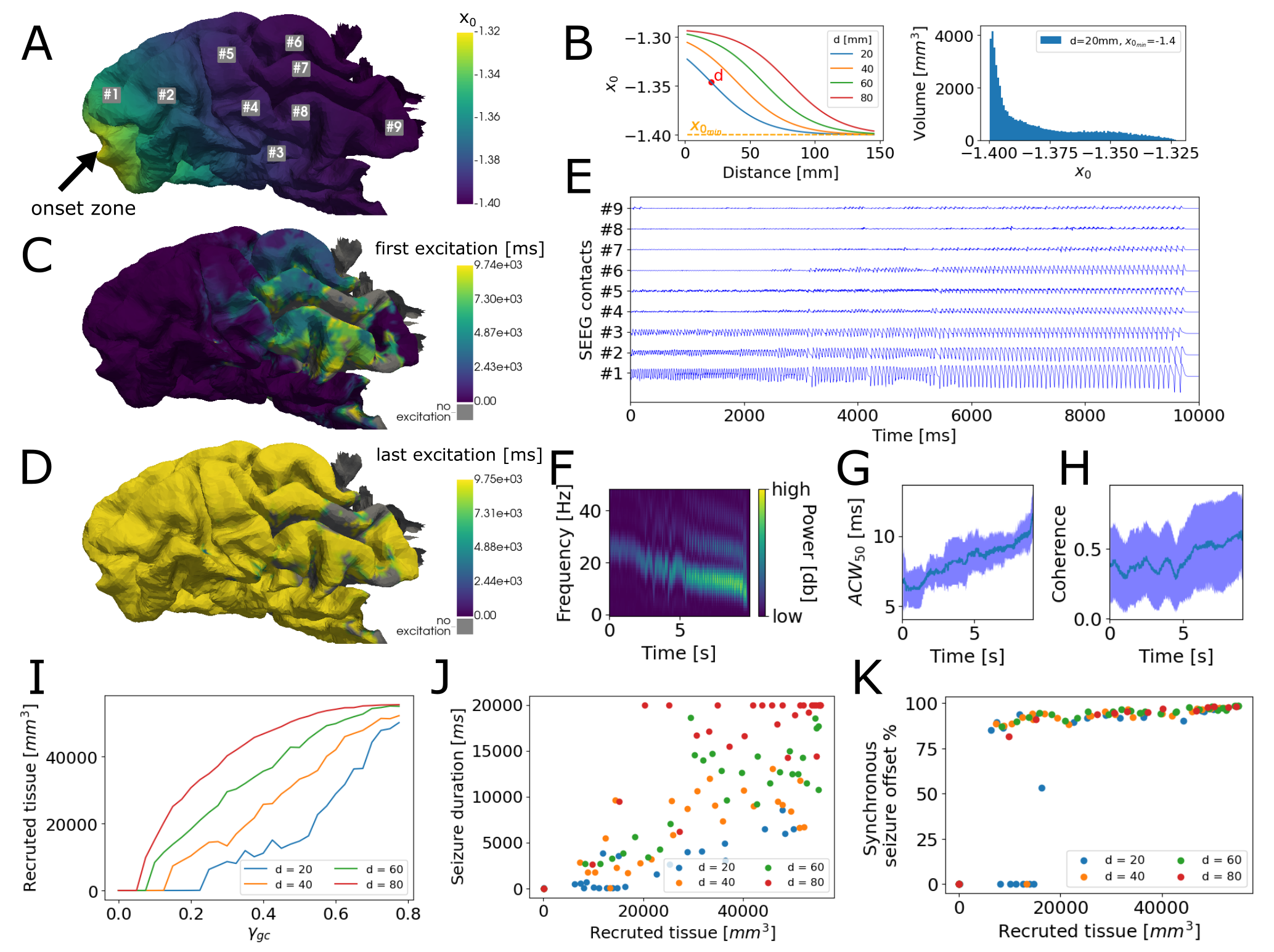}
\centering
\caption{Re-entrant activity invading the propagation zone.}
\textbf{(A)} Lateral view of the left temporal lobe used as EZ for simulation. Color indicates the value of excitability ($x_0$) across the tissue. Numbers indicate the positions of virtually implanted contacts. \textbf{(B)} Left, the sigmoid functions define the decay of $x_0$ with increasing distance from the onset zone on the tip of the anterior temporal pole. Parameters $d$ and $x_{0_{min}}$ indicate the midpoint and the lower horizontal asymptote of the sigmoid function, respectively. Right, histogram of $x_0$ values across the tissue for $d=20mm$ and $x_{0_{min}}=-1.40$. \textbf{(C)} Example simulation using the Epileptor with activity dependent slowing for parameters $d=20mm$, $x_{0_{min}}=-1.40$ and $\gamma_{gc}=0.425$. Color indicates the timing of the first excitation across the tissue. Grey parts of the tissue were not reached by propagation. \textbf{(D)} Same as (C), but color indicates the timing of the last excitation before seizure end. \textbf{(E)} SEEG signals recorded from the simulation in (C) on the contacts displayed in (A). \textbf{(F)} Time frequency plot of the average SEEG signal from (E). \textbf{(G)} Mean (black) and +/- standard deviation (blue) of the evolution of the autocorrelation of the SEEG signals from (E). Precisely, $ACW_{50}$ indicates the lag in ms at which the autocorrelation decays below 0.50. \textbf{(H)} Mean (black) and +/- standard deviation (blue) of the evolution of the pairwise coherence between the SEEG signals from (E).
\textbf{(I)}  Parameter space exploration across coupling strength $\gamma_{gc}$ for $x_{0_{min}}=-1.45$ and different values of $d$. The y-axis indicates the amount of tissue that was excited at least once during the simulation. \textbf{(J)} Scatter plot of seizure duration as a function of the amount of recruited tissue by re-entry activity for the simulations from (I). \textbf{(K)} Scatter plot showing the percentage of recruited tissue terminating re-entry synchronously as a function of the amount of recruited tissue for the simulations from (I).
\label{fig:propagation}
\end{figure}

To examine whether a re-entrant loop can account for the rich propagation repertoire seen in focal seizures, we enlarged the virtual brain to include the entire posterior temporal lobe—Hippocampus, Parahippocampal cortex, Collateral sulcus, Fusiform gyrus, Occipito-temporal sulcus, posterior portions of T3, ITS, T2, STS, T1 (lateral) and T1 (planum temporale), plus the gyrus of Heschl (Fig. \ref{fig:propagation}A). Excitability was distributed as a smooth anterior–posterior gradient: the Epileptor offset $x_0$ followed a sigmoidal decay with distance from the onset zone (Fig. \ref{fig:propagation}B). Thus the anterior temporal lobe formed a highly excitable EZ, while posterior tissue constituted a progressively less excitable propagation zone (PZ).

Empirical recordings show that seizure frequency falls as termination approaches \cite{kramer_human_2012,Jirsa2014}. To reproduce this hallmark we allowed the intrinsic time-scale 
$\tau$  to grow with recent local activity; sustained firing therefore slows subsequent cycles until the loop can no longer re-excite upstream tissue.
With this set-up we ran 20 s episodes from the large EZ (Figs. \ref{fig:propagation}C–E). Three key features emerged:
1) Propagation lag—posterior temporal lobe vertices were recruited only after multiple cycles; a single wave of excitation could not reach them. 2) Delayed SEEG activation—virtual contacts 1–2 (anterior) oscillated from onset, whereas contacts 6–8 (posterior) joined several seconds later (Fig. \ref{fig:propagation}E). 2)
Synchronous termination—once the oscillation had slowed substantially, activity ceased almost simultaneously across EZ and PZ (Fig. \ref{fig:propagation}D).

The slow-down created a clear rise in signal auto-correlation and channel-to-channel coherence, mirroring human depth-EEG (Figs \ref{fig:propagation}F–H). Supplementary Video V2 shows intermittent spiral waves wandering across the inferior temporal surface during the seizure.

\begin{figure}[h]
\includegraphics[width=\textwidth]{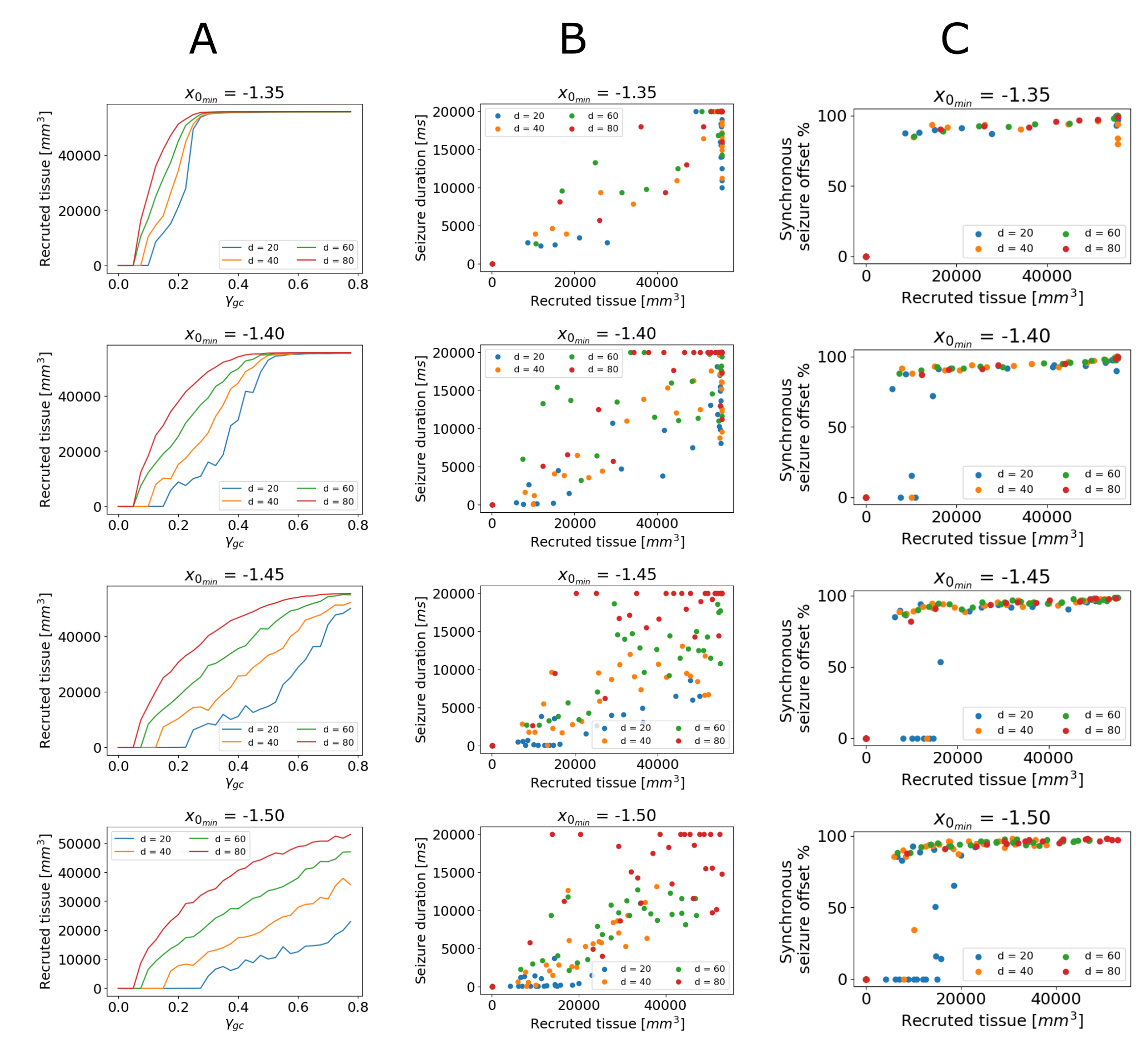}
\centering
\captionsetup{labelformat=empty}
\caption{Figure S7 Parameter exploration for $x_{0_{min}}$ and $d$ for spatial distribution of excitability $x_0$.}
\label{fig:suppl_x0_min}
Each row of scatter plots shows results for a different value of $x_{0_{min}}$. \textbf{(A)} Parameter space exploration across coupling strength. The y-axis indicates the amount of tissue that was excited at least once during the simulation. \textbf{(B)} Scatter plot of seizure duration as a function of the amount of recruted tissue by re-entry activity for the simulations from (A). \textbf{(C)} Scatter plot showing the percentage of recruted tissue terminating re-entry synchronously as a function of the amount of recruted tissue for the simulations from (A).
\end{figure}

Increasing global coupling or shifting the sigmoid midpoint $d$
 posteriorly enlarged the territory that became active (Fig. \ref{fig:propagation}I). Seizure duration was positively correlated with recruited tissue (Fig. \ref{fig:propagation}J); the longest events continued beyond the 20 s window. Filtering for episodes that did terminate revealed a dominant pattern of near-simultaneous offset over wide cortical areas (Fig. \ref{fig:propagation}K). Varying the lower asymptote 
$x_{0_{min}}$, which controls baseline excitability in the posterior lobe, yielded qualitatively identical behaviour (Fig. \ref{fig:suppl_x0_min}).

Together, these results show that a delay-supported re-entrant loop can invade less excitable tissue, decelerate through activity-dependent fatigue, and still shut down synchronously across distant areas — capturing the main spatial and temporal signatures of human focal seizures.

\subsection{Validation in empirical data}

To test the model predictions in vivo, we analysed 184 seizures recorded from 50 drug-resistant epilepsy patients who underwent stereo-EEG. Each recording included a minimum of six depth electrodes and covered temporal and extra-temporal lobe structures. 

\begin{figure}[h]
\includegraphics[width=\textwidth]{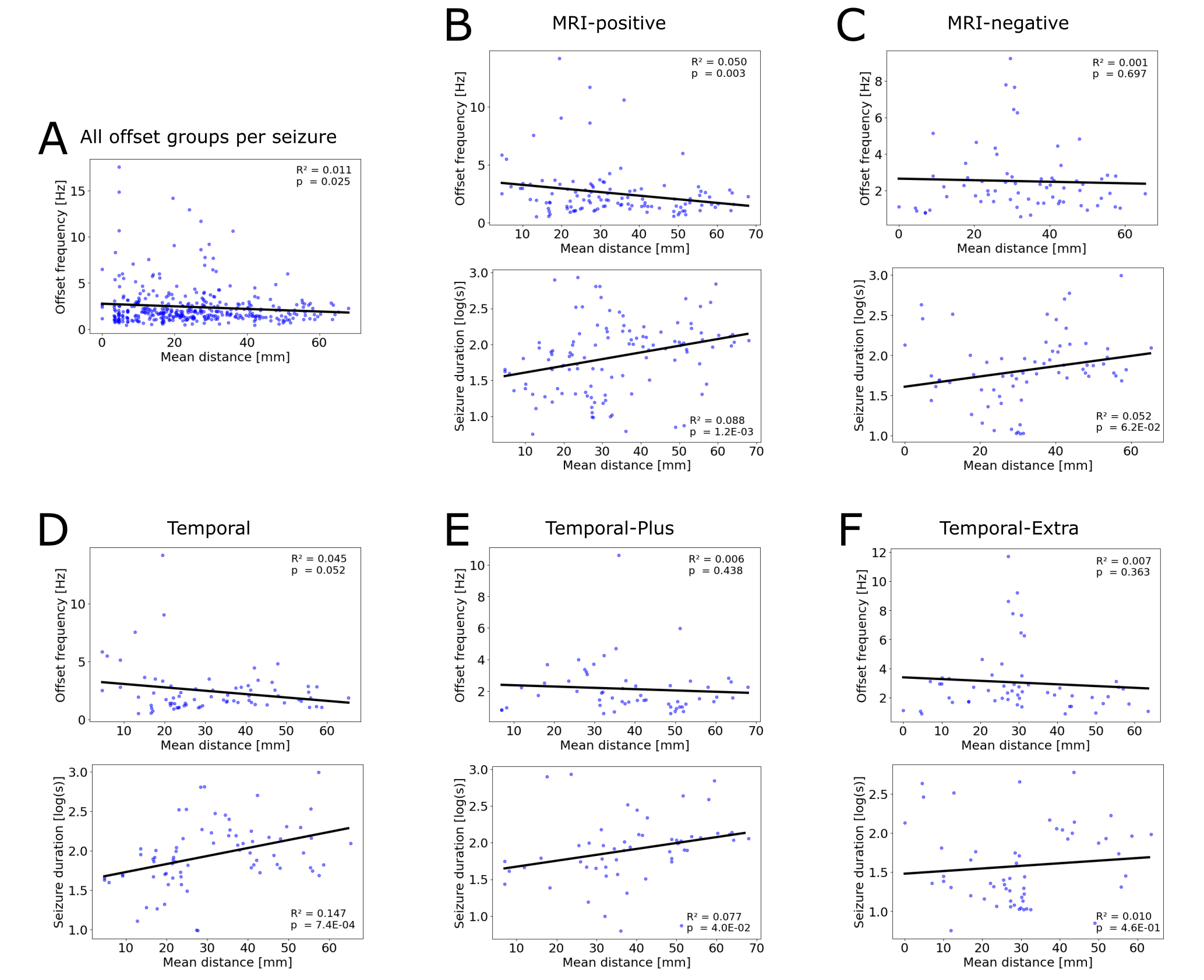}
\centering
\captionsetup{labelformat=empty}
\caption{Figure S8 Empirical data analysis for offset frequency and seizure duration for different conditions.}
\label{fig:suppl_empirical_data}
 \textbf{(A)} Scatter plot of averaged offset frequencies against mean pairwise distance of seizing SEEG contacts. Each dot represents the frequency and mean distance of one synchronous offset group of SEEG channels (n=379). If a seizure has multiple synchronous offsets it can be represented by multiple dots in this plot. The black line indicates a linear model fit with p-value and $R^2$ indicated in the plot. \textbf{(B)} Scatter plot of averaged offset frequencies and seizure durations against mean pairwise distance of seizing SEEG contacts only displaying data from MRI-positive patients (n=116). Black line indicates a linear model fit with p-value and $R^2$ indicated in the plot. \textbf{(C)} Same a (B) but for data from MRI-negative patients (n=68). \textbf{(D)} Same a (B) but for data from patients with Temporal EZ (n=74). \textbf{(E)} Same a (B) but for data from patients with Temporal-Plus EZ (n=55). \textbf{(F)} Same a (B) but for data from patients with Extra-Temporal EZ (n=55).
\end{figure}

For every seizure the offset in each channel was marked visually. Channels whose offsets fell within 1 s were grouped as \emph{synchronous}. When a seizure terminated at different moments across the brain, we retained the largest synchronous group for the primary analysis (Extended Data Fig. \ref{fig:suppl_empirical_data} shows that using all groups yields the same conclusions). Offset frequency was taken as the dominant spectral peak during the final 5 s of ictal activity in that group. To approximate the spatial extent—and thus the intrinsic delay—of the seizing network, we computed the mean Euclidean distance between all contacts in the chosen group.

\begin{figure}[H]
\includegraphics[width=\textwidth]{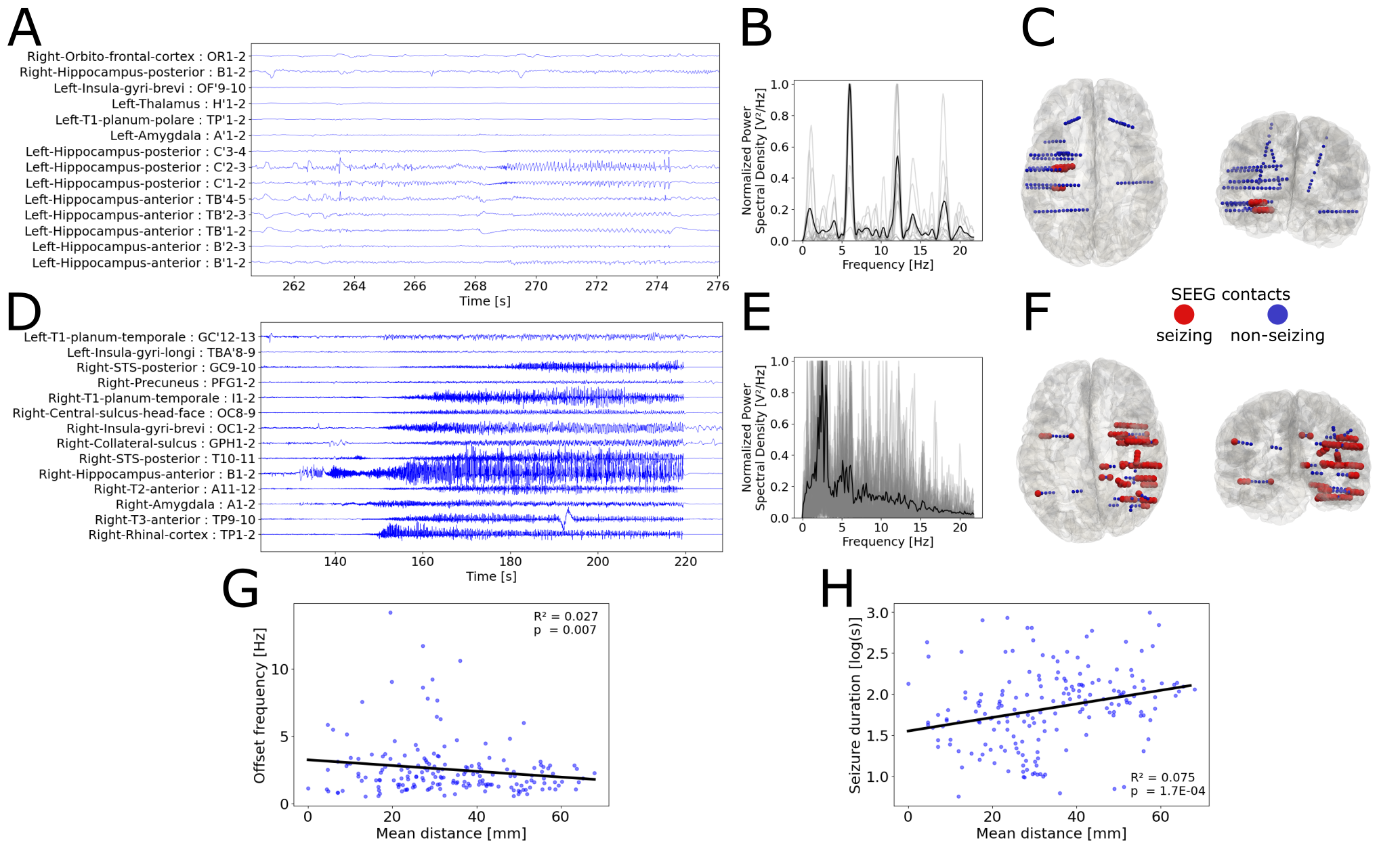}
\centering
\caption{Validation in empirical data}
\textbf{(A)} A subset of SEEG channels in a bipolar montage from a patient with a short (13s) left hippocampal seizure showing synchronous seizure offset. \textbf{(B)} Normalized power spectral density of the last 5s of the seizure of each individual channel (gray) and their average (black) with a peak at around 6Hz. \textbf{(C)} 3D cortical surface plots with the SEEG implantation showing the extent of the seizing tissue. \textbf{(D,E,F)} Same as (A,B,C) but for a different patient with a long (88s) and spatially extended seizure and offset frequency at around 2.5Hz. \textbf{(G)} Scatter plot of averaged offset frequencies against mean pairwise distance of seizing SEEG contacts. Black line indicates a linear model fit with p-value and $R^2$ indicated in the plot. The data show a decreasing trend of the offset frequency with increasing seizing tissue size. \textbf{(H)} Scatter plot of seizure duration against mean pairwise distance of seizing SEEG contacts. Black line indicates a linear model fit with p-value and $R^2$ indicated in the plot. The data show an increasing trend in seizure duration with increasing seizing tissue size.
\label{fig:validation}
\end{figure}

Two representative cases illustrate the extremes (Fig \ref{fig:validation}A-F). A focal seizure restricted to three closely spaced electrodes terminated at 6 Hz after 13 s, whereas a widely distributed seizure spanning eleven electrodes ended at 2.5 Hz after 88 s. These examples mirror the model result that \emph{small loops end quickly and at higher frequencies, large loops linger and slow}. Across the full data set, offset frequency declined 
($R^2$=0.027, p=0.007, Fig \ref{fig:validation}G) and seizure duration rose ($R^2$=0.075, p=1.7E-04, Fig \ref{fig:validation}H) when the mean inter-contact distance increased as predicted by the model in Figs \ref{fig:re-entry_EZ_size}D and \ref{fig:propagation}J.

Repeating the analysis for all synchronous groups confirmed both trends (Extended Data Fig. \ref{fig:suppl_empirical_data} A). Partitioning by structural imaging (MRI-positive vs MRI-negative) or by EZ extent (temporal, temporal-plus, temporal-extra) preserved the duration–size correlation in every subgroup (Extended Data Fig. \ref{fig:suppl_empirical_data}B-F). The offset-frequency trend held in all subgroups but reached statistical significance only in MRI-positive patients, possibly reflecting better delineation of the active network.

Taken together, empirical seizures obey the same inverse relationship between loop size, termination frequency and duration that emerges from the high-resolution virtual brain, supporting delay-mediated re-entry as a unifying mechanism across patients.

\section{Discussion}

We introduced a patient-specific virtual brain that reproduces re-entrant seizure dynamics at millimetre resolution. By placing the Epileptor in an excitable regime on cortex and sub-cortex, a modest perturbation was sufficient to ignite a loop in the temporal lobe; its persistence hinged on coupling strength, axonal delay and the spatial extent of the epileptogenic zone (EZ). Parameter sweeps revealed three canonical behaviours: sustained re-entry, self-limiting transients and fast wave fronts that extinguish once all tissue is refractory. Phase-targeted biphasic stimulation aborted re-entry within a narrow timing window, whereas local or global virtual lesions either stabilised, destabilised or reshaped the circulating wave. The same model reproduced activity-dependent slowing and synchronous seizure termination—phenomena repeatedly documented in intracranial EEG.

This work is, to our knowledge, the first to use sub-millimetre diffusion-MRI connectomes with spatially continuous neural fields in epilepsy. Vertex-wise connectomes have illuminated intra-areal topology, and neural-field studies have separately captured seizure onset and wave propagation on simplified geometries. Only a deep-brain-stimulation simulation has previously combined a high-resolution connectome with dynamical units, and that omitted intra-cortical connectivity\cite{an_high-resolution_2022}. Our integration therefore represents a substantive step toward a bona fide digital twin for surgical planning.

Re-entrant excitation has long been observed in hippocampal slices, where recurrent mossy-fibre sprouting sustains self-propagating bursts\cite{sutula_unmasking_2007}. Modelling studies that turn select dentate granule cells into hubs reproduce those after-discharges, hinting that micro-circuit rewiring can seed macro-scale loops \cite{morgan_nonrandom_2008}. Multi-scale co-simulation now makes it feasible to embed such cellular detail inside whole-brain fields, an attractive route for dissecting how local plasticity translates into network-level seizures\cite{kusch_multiscale_2024}.

White-matter delays proved to be the key moderator of re-entry. Too little coupling and waves fizzled; too much coupling triggered rapid, global refractoriness; only an intermediate corridor of delays and gains produced loops that matched empirical seizure durations. Within that corridor, propagation slowed as tissue fatigued, eventually dropping the loop frequency below the delay-supported threshold and forcing collapse—an intrinsic self-termination mechanism that requires no slow variable or bistability. Single-pulse after-discharges seen during presurgical mapping, and their reproduction in spatially extended Jansen–Rit models, likely reflect a similar interaction between delay and refractoriness \cite{Goodfellow2012}.

Our stimulation experiments align with rodent data showing that single or few phase-locked pulses, rather than high-frequency trains, can terminate seizures if precisely timed\cite{
takeuchi_closed-loop_2021,osorio_seizure_2009}. Charge-balanced waveforms mitigate the cytotoxic risk inherent to DC pulses and may therefore be more translatable. Inclusion of thalamic circuits will refine phase targets for spike–wave discharges, where cortico-thalamic loops dominate.

Structural interventions likewise showed promise. Simulated radiofrequency thermocoagulation of white-matter pathways altered re-entry dynamics in ways consistent with emerging ideas of tract-focused epilepsy surgery. Multiple subpial transection (MTS) slowed or abolished loops, mirroring animal data\cite{sugiyama_electrophysiological_1995, hashizume_multiple_1998}. In current clinical applications, SEEG-RFTC is mainly used to target gray matter malformation, like cortical dysplasia or heterotopia \cite{catenoix_seeg-guided_2008} and not white matter pathways. However, stimulation of white matter targets to disrupt epileptic networks has already been suggested\cite{girgis_white_2016}. The lesioning of pathways to prevent seizure spread has been  practiced only in lobe-disconnection\cite{kishima_which_2013} and MTS\cite{rolston_multiple_2018}. A variant of MTS is the hippocampal transection which can reduce seizures, with less side effects\cite{shimizu_hippocampal_2006}. These results support the concept that selectively disabling a handful of well-chosen long-range fibres may rival larger cortical resections, a process, which can be guided by the virtual brain on an individual level. 

Parallels with cardiac arrhythmia are striking : both heart and brain are delay-coupled excitable networks \cite{trayanova_whole-heart_2011,Tse2016} in which spiral-wave re-entry responds to ablation or phase-resetting shocks \cite{kingma_overview_2023}. Reflection of travelling waves at heterogeneities — documented at the V1/V2 boundary \cite{xu_compression_2007} — could further enrich the repertoire of cortical re-entry modes and may surface in our model under stronger propagation-dependent slowing.

The extended EZ also projected diverse onset–offset signatures onto virtual SEEG. Sites near the trigger resembled a saddle-node bifurcation at onset, whereas propagation zones displayed supercritical Hopf-like growth; all channels converged onto homoclinic-looking offsets. These patterns emerge without local bifurcations, underscoring how network dynamics can mimic classical critical transitions on sensors \cite{sip_evidence_2021}. Activity-dependent slowing increased inter-channel synchrony and autocorrelation, again matching patient recordings.

Several limitations warrant mention. The model has not yet been fitted to individual electrophysiology; spherical-harmonic compression may enable future parameter inference across tens of thousands of vertices \cite{daini_spherical-harmonics_2020}. Tractography biases remain \cite{zhang_quantitative_2022}, although forthcoming ultra-high-field post-mortem datasets will offer calibration ground truth \cite{Beaujoin2020}. Finally, we explored only one or two onset/offset archetypes, whereas broader taxonomies of seizure types almost certainly implicate additional network motifs \cite{saggio_taxonomy_2020}.

These encouraging in-silico results support a first-in-human, phase-targeted single-pulse-stimulation trial guided by high-resolution virtual-brain modelling.

\section{Methods}
\subsection{Patient characteristics and data acquisition}
We analyzed data from 50 patients with drug-resistant focal epilepsy who underwent a standard presurgical examination at La Timone Hospital in Marseille, France. Informed written consent was obtained from all patients in accordance with the Declaration of Helsinki, and the study protocol was approved by the local ethics committee (Comité de Protection des Personnes Sud-Méditerranée 1). Each patient underwent a comprehensive presurgical evaluation, including medical history assessment,  neurological examinations, neuropsychological evaluation, fluorodeoxyglucose PET, high-resolution 3T MRI, long-term scalp EEG monitoring, and invasive SEEG recordings.

Data from a single patient (female, 35y) was selected to build a high resolution virtual brain model and simulate re-entry dynamics. This patient underwent SEEG implantation after which the EZ was estimated to extend across the left Hippocampus, Amygdala, Temporal pole, Rhinal and Parahippocampal cortex. Left anterior temporal lobe resection was performed with a good outcome (Engel Score 2). During pre-surgical evaluation a T1 weighted image (sagittal MPRAGE sequence, $2.98 ms/2.3s$ (TE/TR), voxel size $1.0 \times 1.0 \times 1.0  mm$, image dimensions $160 \times 256 \times 256$ voxels) and a diffusion weighted scan ($88ms/3s$ (TE/TR), voxel size $2.0 \times 2.0 \times 2.0 mm$, image dimension $126 \times 114 \times 69$ voxels, bval=$[0,1000,1400,1800]$, with an angular gradient set of 193 directions) was acquired on a 3T Siemens Verio MRI scanner.

\subsection{Structural scaffold construction}
Using neuro-imaging data we build a high resolution virtual brain model of the patient. The T1w MR image is processed using the recon-all pipeline of the Freesufer v7.1.1 software package \cite{Fischl2012} to perform volume segmentation and cortical surface reconstruction. The VEP atlas \cite{Wang2021} was mapped onto the cortical surface of the subject to obtain a regional parcellation of the brain. Additionally, the T1w MRI image was processed with the HippUnfold BIDS app \cite{Dekraker2021} in order to obtain a further detailed segmentation and surfaces of the hippocampal formation (subiculum, CA1-4 and the dentate gyrus). The cortical and hippocampal surfaces were connected to create an anatomical realistic surface, in which the hippocampal surface represents the continuation of neocortex \cite{Duvernoy1988}.  First, the part of the cortical surface representing the medial wall was removed. Second, a triangulation was inserted connecting the boundary of the rhinal and parahippocampal cortex with the boundary of the subiculum of the hippocampal surface. Third, the mesh around the new triangulation was smoothed using Laplacian smoothing, to improve the shape and size of triangles.
The surface of the cerebellum was created using the surface reconstruction stream of Freesurfer's recon-all pipeline and FSL's FAST algorithm \cite{Zhang2001}. First, a cerebellar mask was created from the Freesurfer white and gray matter cerebellum segmentation and it was used to extract the cerebellum of the T1w image. Second, FAST was used on the T1w cerebellum to obtain a detailed white and gray matter segmentation. Third, the algorithms of the Freesurfer surface reconstruction pipeline were used, to tesselate a surface around the cerebellar white matter, fix its topology and project it onto the gray matter - cerebrospinal fluid boundary of the cerebellum. 
Surfaces are a natural representation of the hippocampus and the cerebellum. Other subcortical structures however can not be represented by a surface, because they fill a volume with multiple nuclei of cell bodies. Therefore we chose a point grid filling the volume of thalamus, caudate nucleus, putamen, pallidum, amygdala and nucleus accumbens, as their representation in the high-resolution model. These structures are readily segmented by the recon-all pipeline. However, we would like to represent volumetric structures at a comparable resolution to surfacic structures. Each vertex of the cortical surface mesh, does not only represent an area of the surface, but also a volume of the corresponding cortical gray matter ribbon. In order to assign a comparable volume to each volumetric grid point we calculated the area that each cortical vertex represents. Because the pial surface is a projection of the white matter surface along the voxel gray level intensities, the vertices and triangles of the two surfaces correspond to each other. Thus, a natural way to compute the volume of a vertex, is to compute the volume of the "triangular prism"-like shape that results from the projection of a triangle from white to pial surface and add a third of this volume to each corner vertex. Since this shape is not an exact triangular prism, but a truncated, rotated and shifted version, we use numeric integration to estimate its volume. The "triangular prism"-like shape is subdivided by projecting the white surface triangle towards its corresponding pial surface triangle in 100 steps, resulting in 101 triangular prism like subvolumes. Each subvolume is computed by using a convex hull approach. Finally, the mean vertex volume across the whole cortical surface was used to resample the subcortical segmentation to an equal isotropic voxel size (=1.77$mm^3$).

\subsection{Connectivity estimation}
To assess structural connectivity between cortical and subcortical vertices we processed the diffusion imaging data using the MRtrix3 software package \cite{Tournier2019}. First, the diffusion data was denoised \cite{Veraart2016} and Gibbs ringing artifact corrected \cite{Kellner2016}. Second, movement, epi-distortion and eddy current correction was performed \cite{Andersson2016}. Because the diffusion data was acquired in anterior to posterior phase encoding direction only, we used Synb0 DisCo \cite{Schilling2019} to estimate a synthetic undistorted b0 image that was used for distortion correction. The algorithm from \cite{Dhollander2019} was used to estimate response functions for gray matter, white matter and cerebrospinal fluid diffusion signal. Subsequently, fiber orientation distributions (FOD) were estimated per voxel using multi-shell multi-tissue constrained spherical deconvolution \cite{Dhollander2019} and corrected for intensity inhomogeneities \cite{Raffelt2017}. 10M streamlines were sampled using probabilistic tractography (iFOD2) \cite{Tournier2010} with anatomical constraints \cite{Smith2012}. Streamlines were weighted according to the SIFT2 algorithm \cite{Smith2015}, to match the streamline densities with the underlying fiber densities as estimated through spherical deconvolution. Anatomical constraints were given by the Freesurfer anatomical segmentation, in which we substituted the hippocampal and cerebellar volume by the segmentation from HippUnfold and FSL's FAST, respectively. 
Due to the probabilistic nature of tractography spurious tracts can be generated. We noticed this primarily when constraining our simulation to the EZ of the subjects. Tracts would start within the anterior temporal lobe, travel towards the occipital lobe and return. These few tracts would cause long delays in the system and would overestimate the re-entry effect. Therefore we thresholded the tracts connecting within the EZ to the 95th percentile, corresponding to a maximum tract length of 42.87mm. Any longer tracts were removed. We visually confirmed that the remaining fibers would reside inside the volume of the EZ.
Subsequently, a vertex to vertex based connectome was constructed. For each streamline a radial search was performed at both end points to find a subcortical grid point that it can be assigned to. The search is performed within a radius of subcortical voxel edge length  (1.21mm) + 0.01 mm. If no subcortical grid point is found, the Möller-Trumbore ray triangle intersection algorithm \cite{Moller2005} is used to find an intersection of the ray, that is given by the last 2 points of the streamline, with a triangle of the cortical mesh. A triangle of the mesh was only checked for intersection, if all 3 vertices of the triangle were on average not further distant to the ray origin than 5mm. If an intersection of the ray and a triangle was found, the streamline was assigned to be connected to the vertex of the triangle that is closest to the intersection point. If one or both of the streamline ends could not be assigned to either a subcortical grid point or to a cortical vertex, the tract was ignored and not used to construct the connectome. To compute connection weights and lengths, we used the sum of SIFT2 weights and the average lengths of all connecting streamlines between two vertices, respectively.
Next to long range connections through white matter fibers, our model accounts for distance based local coupling between neighboring vertices on the surface or the volumetric grid. Therefore a vertex to vertex distance matrix is computed. We use geodesic and euclidean  distances for surfaces and subcortical grids, respectively. Distances are only computed between vertices of the same surface or subcortical grid region, because they are otherwise topologically disconnected. An exception to this is the connection between hippocampal structures CA3/CA4 and the dentate gyrus. Even if the surfaces are topologically disconnected, both structures are in close spatial proximity and are connected via local connections \cite{Nastenko2021}. We therefore added to the distance matrix the euclidean distances between boundary vertices belonging to the CA3/CA4 part of the hippocampal surface and the vertices of the dentate gyrus surface. Distances were only computed between vertices which are in the same range along the anterior-posterior axis, given the unfolded space from HippUnfold.

\subsection{Model simulation}
To investigate seizure like dynamics with re-entrant excitation in the high resolution virtual brain model we equip the structural scaffold with a dynamical neural field model. The Epileptor \cite{Jirsa2014} is a phenomenological neural mass model which describes seizure onset, propagation and offset and which was recently extended to neural fields \cite{Proix2018}. Using averaging methods and exploiting time scale separation the Epileptor can be reduced to a two dimensional dynamical system, the 2D Epileptor \cite{Proix2014}. The system of differential equations that governs the dynamics of every vertex $i$ in the brain network is given by : 
\begin{equation}
\renewcommand*{\arraystretch}{1.5}
\label{eq:epileptor2D}
    \begin{array}{lll}
        \dot{x_i}(t) =& \frac{1}{\tau} (-x_i(t)^3 -2x_i(t)^2-z(t)+I \\
            &+ \gamma_{gc} \sum_{j=0}^{n} w_{i,j} H(x_j(t-\frac{d_{i,j}}{v_{gc}}),\theta) \\
            &+ \gamma_{lc}     \sum_{j=0}^{n} V_j L(g_{i,j}) H(x_j(t-\frac{g_{i,j}}{v_{lc}}),\theta) + I_{ext}(t)) \\
        \dot{z_i}(t) =& \frac{1}{\tau}( \frac{\epsilon}{x_i(t)^2+1} (4(x_i(t)-x_0)-z_i(t)))   
    \end{array}
\end{equation}
with 
\begin{equation}
    \begin{array}{lll}
        H(x,\theta) &= 
        
            \begin{cases}
                1, & \text{if } x\geq \theta\\
                0, & \text{otherwise}
            \end{cases} \\
        L(g)  &= \frac{1}{2}e^{-|g|} 
    \end{array}
\end{equation}
$x$ and $z$ represent a fast and slow state variable, respectively, describing the switching between ictal and inter-ictal states. However, for our purpose we will use it to describe oscillatory activity during seizures. We slightly modified the equation from the original 2D Epileptor \cite{Proix2014} by dividing $\epsilon$ by $(x_i^2+1)$ in the equation for $\dot{z_i}$. 
This was done to bias the duration of the oscillatory cycle towards the refractory period, as has been observed empirically that the spike-width is shorter than the inter-spike interval \cite{smith_ictal_2016}. This modification does not change the original bifurcation structure and phase plane of the mode. It simply scales the derivative differently across different parts of the phase plane.
Each Epileptor generates outputs and receives inputs to and from the other $n-1$ vertices of the network. These interactions are global or local. Global connections between vertices $i$ and $j$, through white matter tracts, are scaled by $\gamma_{gc}$, weighted by the corresponding element $w_{i,j}$ of the connectivity matrix and delayed by the fraction of tract length $d_{i,j}$ over global conduction speed $v_{gc}$. 
A Heaviside step function $H(x,\theta)$ with threshold $\theta$ is used for coupling Epileptors and can be interpreted as a firing rate function, i.e. only when an Epileptor is seizing it would project to other nodes of the network. Local connection weights, based on the geodesic or euclidean distance $g_{i,j}$ between neighboring vertices of the cortical mesh or subcortical grid, respectively, are computed using a Laplacian kernel $L(g)$. 
In the continuous case, adding up the local connections corresponds to calculating a surface integral for cortical surfaces or a volume integral for subcortical volumes. To enable comparison, we assign a volume to the cortical surface by taking into consideration the thickness between white and pial surface. 
Thus each vertex of the network is weighted by its volume $V_j$. Finally, external stimulation can be applied to the model via $I_{ext}(t)$. We use following values for the parameters if not stated otherwise in the results section $\epsilon=0.015, x_0=-1.2916, I=1, \theta=0.2, v_{gc}=3.9mm/ms$ \cite{lemarechal_brain_2022}$, v_{lc}=0.33mm/ms $ \cite{salami_change_2003}. 

To mimic the slowing in frequency during seizures, we implemented an activity dependent slowing mechanism into the model. The parameter $\tau$ becomes the state variable $\tau_i(t)$ for each Epileptor and the full dynamical system becomes :
\begin{equation}
\renewcommand*{\arraystretch}{1.5}
\label{eq:epileptor2D_slowing}
    \begin{array}{lll}
        \dot{x_i}(t) =& \frac{1}{\tau_i(t)} (-x_i(t)^3 -2x_i(t)^2-z(t)+I \\
            &+ \gamma_{gc} \sum_{j=0}^{n} w_{i,j} H(x_j(t-\frac{d_{i,j}}{v_{gc}}),\theta) \\
            &+ \gamma_{lc}     \sum_{j=0}^{n} V_j L(g_{i,j}) H(x_j(t-\frac{g_{i,j}}{v_{lc}}),\theta) + I_{ext}(t)) \\
        \dot{z_i}(t) =& \frac{1}{\tau_i(t)}( \frac{\epsilon}{x_i(t)^2+1} (4(x_i(t)-x_0)-z_i(t)))    \\
     \end{array}
\end{equation}
For which we define $\mu$ as the inverse of the timescale $\mu_i(t) = \frac{1}{\tau_i(t)}$, which evolves in time as :
\begin{equation}
\renewcommand*{\arraystretch}{1.5}
\label{eq:epileptor2D_slowing}
    \begin{array}{lll}
        \dot{\mu_i}(t) =& 
        \begin{cases}
            -\mu_{nmax} * (\mu_i(t)-\mu_{nc}), & \text{if } x_i(t)\leq 0\\
            -\mu_{pmax} * (\mu_i(t)-\mu_{pc}), & \text{otherwise}
        \end{cases} \\
    \end{array}
\end{equation}
The dynamics of $\mu_i(t)$ are controlled by two linear differential equations, respectively increasing and decreasing $\mu$ towards $\mu_{nc}$ and $\mu_{pc}$ whenever the system is in its refractory ($x_i(t)\leq0$) or active period otherwise. $\mu_i(0)=8$ is initialized sufficiently high to have fast oscillations at the beginning of the seizure and decays with ongoing re-entry activity towards $\mu_{pc}=1$ with $\mu_{pmax}=0.002$ controlling the rate of decay. The rate of increase $\mu_{nmax}=0.000004$ of $\mu_i(t)$ towards $\mu_{nc}=8$ is chosen sufficiently slow, such that during ongoing re-entry activity the system is guaranteed to slow down.
The system of differential equations is solved numerically using a deterministic Heun integration scheme with a step size of 0.01.

\subsection{SEEG electromagnetic forward model}
Mapping of neural activity onto SEEG sensors is done through solving the electromagnetic forward problem. We assume that the origin of extracellular fields, measurable with SEEG, is mainly attributed to electrical dipoles stemming from post-synaptic currents of pyramidal cells which are oriented normal to the cortical surface \cite{Buzsaki2012}.  Location and orientation of the dipoles is therefore given by the midthickness cortical surface and its surface normal. For subcortical structures, which are modeled as volumetric grids, we assume a random dipole orientation at each grid point. We neglect possible boundary effects, assuming SEEG contacts are sufficiently close to the electric source and far from tissue boundaries with different electrical conductivity, and use the analytical solution for local field potentials in an unbounded homogeneous medium from Sarvas et al.\cite{Sarvas1987}. The electric potential $P$ that is recorded at SEEG contact $k$ due to neural activity at vertex $i$ is given by
\begin{equation}
    P_{i,k}=\frac{a_i}{4\pi\sigma}Q* \frac{r_k-r_i}{|r_k-r_i|^3}
\end{equation}
where $r_k$ and $r_i$ are the position vectors of contact $k$ and source vertex $i$ respectively.  $|v| $ represents the L2 norm of a vector $v$. $Q$ is the dipole orientation vector and $\sigma$ the electrical conductivity. $a_i$ represents the volume associated with a given source vertex $i$. Since we assume constant conductivity across the brain $\sigma$ becomes merely a scaling factor which we set to $\sigma = 1$.

\section{Data availability}
The data that support the findings of this study are not publicly available due to privacy restrictions. Access to the data may be granted upon reasonable request.

\section{Code availability}
All code to process brain imaging data and reproduce the results reported in this manuscript can be found in the following Ebrains collaboratory which will be made public upon publication \\
(https://wiki.ebrains.eu/bin/view/Collabs/reentry-seizure-model).

\section{Acknowledgments}
This project has received funding from the European Union’s Horizon Europe Programme under the Specific Grant Agreement No. 101147319 (EBRAINS 2.0 Project).

\section{Author contributions}
Conceptualization, P.T., H.W. and V.J.; methodology, P.T., H.W. and V.J.; software P.T. and M.W.; writing-original draft, P.T. and V.J.; writing - review and editing, P.T., H.W., M.G., F.B. and V.J.; resources, M.G. and F.B.; funding acquisition, V.J.; supervision, H.W. and V.J.

\section{Competing interests} 
The authors declare no competing interests.

\printbibliography

\section{Supplementary Information}
\beginsupplement

\begin{figure}[h]
\includegraphics[width=\textwidth]{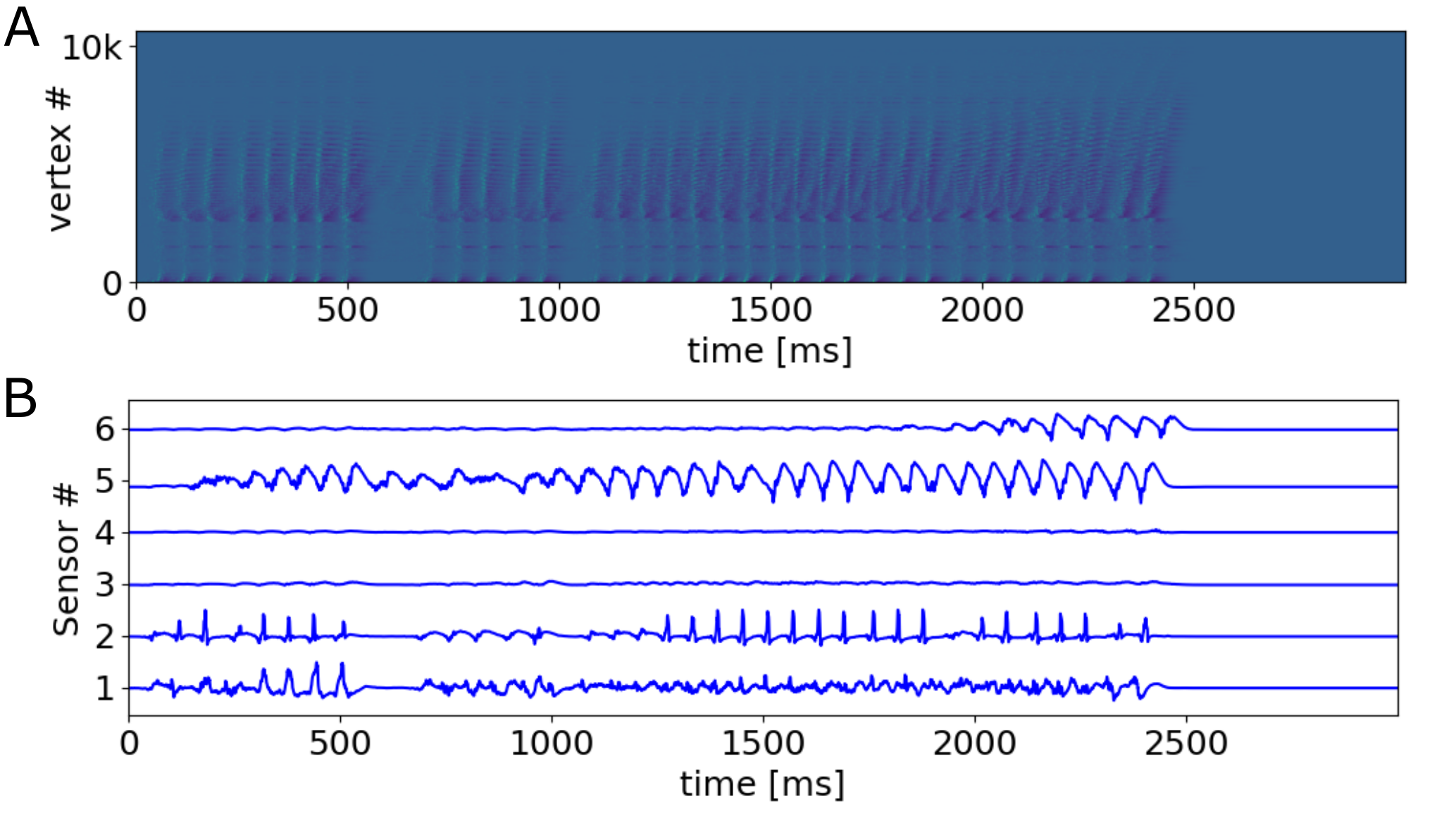}
\centering
\captionsetup{labelformat=empty}
\caption{Figure S2 Self limiting re-entry excitation in a patch of the cortex.}
\label{fig:suppl_self_limiting_re-entry}
\textbf{(A)} Space-time plot, showing re-entrant neural activity. \textbf{(B)} Electrical signal on the contacts of (Fig \ref{fig:re-entry_temporal_lobe} C) given the simulation in (A).
\end{figure}

\end{document}